\documentclass[12pt,aps,prl,amssymb,amsmath,superscriptaddress,showpacs,floatfix,reprint]{revtex4-2}
\usepackage{mathtools}
\usepackage{epsfig}
\usepackage{graphicx}
\usepackage[english]{babel}
\usepackage{xcolor} % to make changes visible in drafts, remove it after finalizing
\usepackage{soul}%to make changes visible in drafts,
\usepackage[unicode=true]{hyperref}
\hypersetup{breaklinks=true, colorlinks=true, urlcolor= blue, citecolor=blue}
\begin{document}

\title{Universal Fast Mode and Potential-dependent Regimes in Wetting Kinetics}

%\author{Syed Shuja Hasan Zaidi$^1$, Prabhat K. Jaiswal$^1$, Madhu Priya$^2$, and Sanjay Puri$^3$}
%>\email[Corresponding author:\; ]{}
\author{Syed Shuja Hasan Zaidi}
\author{Prabhat K. Jaiswal}
\email[Corresponding author:\;]{prabhat.jaiswal@iitj.ac.in}
\affiliation{Department of Physics, Indian Institute of Technology Jodhpur, Karwar 342030, India}
\author{Madhu Priya}
%>\email[]{madhupriya@bitmesra.ac.in}
\affiliation{Department of Physics, Birla Institute of Technology Mesra, Ranchi  835215, India}
\author{Sanjay Puri}
\email[Corresponding author:\;]{puri@mail.jnu.ac.in}
\affiliation{School of Physical Sciences, Jawaharlal Nehru University, New Delhi 110067, India}

%\affiliation{$^1$Department of Physics, Indian Institute of Technology Jodhpur, Karwar 342030, India}
%\affiliation{$^2$Department of Physics, Birla Institute of Technology Mesra, Ranchi  835215, India}
%\affiliation{$^3$School of Physical Sciences, Jawaharlal Nehru University, New Delhi 110067, India}

\date{\today}
\begin{abstract}
	We present simulation results from a comprehensive molecular dynamics (MD) study of surface-directed spinodal decomposition (SDSD) in unstable symmetric binary mixtures at wetting surfaces. We consider long-ranged and short-ranged surface fields to investigate the early-stage wetting kinetics. The attractive part of the long-ranged potential is of the form $V(z) \sim z^{-n}$, where $z$ is the distance from the surface and $n$ is the power-law exponent. We find that the wetting-layer thickness $R_1(t)$ at very early times exhibits a power-law growth with an exponent $\alpha = 1/(n+2)$. It then crosses over to a universal fast-mode regime with $\alpha=3/2$. In contrast, for the short-ranged surface potential, a logarithmic behavior in $R_1(t)$ is observed at initial times. Remarkably, similar rapid growth is seen in this case too.  We provide phenomenological arguments to understand these growth laws. Our MD results firmly establish the existence of universal fast-mode kinetics and settle the related controversy.
\end{abstract}

\maketitle

\emph{Introduction.} An immiscible binary ($A+B$) mixture in a homogeneous phase segregates into regions of $A$-rich and $B$-rich domains when quenched deep below the miscibility curve. This spontaneous decay of the unstable mixture is called \emph{spinodal decomposition} (SD). This is an important example of coarsening or domain growth problems, which have received great attention in the literature \cite{VS09,bray02,onuki02}. Multicomponent mixtures and their phase separation have immense scientific and technological applications in complex fluids such as polymer blends \cite{MRA95,RLE91}, emulsions \cite{MJ14,SD11}, alloys and glasses \cite{DEE14,MH21}, Ionic Liquids (ILs) \cite{JFJ09}, active materials for organic solar cells \cite{LEV05} and memristors \cite{NYY21}, hardening of alloys \cite{ZLY21}, stabilization of proteins using ILs \cite{YNH20}, composite materials with enhanced mechanical strength and plasticity \cite{MJH09,TH16,KY19,MP20}.

\par The morphology of domains formed due to phase segregation could be bicontinuous (interconnected) or droplets, depending upon the concentration ratio of the components in the mixture \cite{VS09}. These coarsening domains follow power-law growth as $\ell (t) \sim t^{\theta}$, where the exponent $\theta$ defines the underlying physical mechanism for the transport of particles.  The $\theta=1/3$ for phase-separating binary solid alloys marks the diffusion as the primary mode of transportation in solids and is referred to as Lifshitz-Slyozov (LS) growth law \cite{IV61}. For fluid mixtures, hydrodynamics comes into play at later times, and the resulting evolution exhibits a crossover to other exponent values, e.g., $\theta=1$ followed by $\theta=2/3$ in three-dimension (3D), owing to a faster advective transport of the particles \cite{VS09,bray02,VMI01,AM01,SSS10,SSS12}. 

\par Introducing a substrate ($ S $), with a preferential attraction for one of the components (say, $ A $) of this phase-segregating mixture $A+B$, breaks the translational symmetry of the particle transportation in a direction perpendicular to the surface. It changes the kinetics and domain morphology at the surface \cite{puri05,KSS10}. 
If the system is now quenched below the spinodal curve, it spontaneously decomposes into $ A $-rich and $ B $-rich domains. Simultaneously, the system is wetted by the preferred component $ A $. The wetting phenomenon of the surface in conjunction with the bulk phase separation is known as \emph{surface-directed spinodal decomposition} (SDSD). The surface may evolve either in a \emph{completely wet} (CW) or a \emph{partially wet} (PW) morphology at equilibrium, depending upon the relative surface tensions between $ A, B $, and $ S $. 
 
\par The SDSD phenomenon is effective in fabricating low-cost, lightweight, and easy-to-process optoelectronic devices \cite{JDM15,SMJ97,Michels11}. Phase-separated structures, including bicontinuous interconnected layers, droplets, pillars, and disks in organic thin films, ferroelectric transistors, organic photovoltaic devices (OPVs), etc., can be tailored to achieve distinct optical and electrical properties \cite{LEV05,JRI10,KDB08}. In particular, an OPV consists of an active layer featuring interconnected domain morphology for effective charge separation and layering near the substrate. This gives rise to an enhanced charge collection and improves the efficiency of optoelectronic devices \cite{Chen19,YDL11,Michels11,HYX20,CJP17}. Additionally, layering in SDSD can also enhance the physical, mechanical, and surface properties of polymer blends \cite{MRA95,YDL11}. The layering, however, may also result in undesirable enrichment layers adversely affecting the mechanical and thermal properties of polymer blends \cite{YDL11,MPJ16}. Furthermore, SDSD provides methods for generating transient target morphologies in metallic or polymer mixtures (block copolymers) \cite{TR16}, etc, via chemically or morphologically patterned substrates or through the introduction of micro- or nano-sized filler particles \cite{PPS20,PPS20-2,SAR20}. These target morphologies show potential technological applications in bit-patterned media \cite{CTJ16}, nanowires \cite{PAC15}, polarizers \cite{SJI14}, ion conduction channels \cite{JGH18}, nanolithography \cite{CMD14}, and electrolytes in energy storage devices \cite{TG19}. A few more recent SDSD applications include thin-film fabrication for oral drug delivery \cite{PMA21}, microfluidic-based liquid-liquid phase separation (water-ILs) to recover on demand the IL-rich phase of interest  \cite{MAM21}, and food processing \cite{ERD12}. In the light of the above-mentioned technological applications, it is compelling to understand and control the growth regimes of layered morphologies and coarsening bulk domains in SDSD.

\par Much scientific effort has been put into SDSD through analytical investigations \cite{SK92,troian93prl,troian92,HPW97,RR90,KH91,marko93}, experimental studies \cite{MRA95,RLE91,MG03,krausch95, PA91,APF92,BCA93,Michels11,YDL11}, and computer simulations \cite{SY88,SH97,puri05,binder83,GA92,SK01,SK94,PSS12,SSJ01,SSJ05,SSJ06-pre,bray02,APF21,toxvaerd99prl,HA97pre,PWA94prl,HT00,tanaka01,SAR20,PPS20,PPS20-2,MPJ16}. The Puri-Binder model (PB) \cite{SK92} is the first successful theoretical model of SDSD for binary alloys and early-stage kinetics of polymer mixtures at the coarse-grained scale. 
This model consists of Cahn-Hilliard-Cook (CHC) equation for the phase-separating mixture supplemented by two boundary conditions at the surface. Puri and Binder studied the kinetics of SDSD of critical mixtures ($50\%A - 50\%B$) for a short-ranged $ \delta $-function potential and formulated tools to characterize domain growth parallel and perpendicular to the wetting surface \cite{SK92-1}. They found that the growth of the wetting-layer thickness $R_1(t)$ appeared to be logarithmic with time $ t $. 

\par Puri and Binder \cite{SK01,SK94} considered the case when the surface exerts a long-ranged attractive force on the preferred component of the mixture. For surface interactions of the form $ V(z) \sim z^{-n} $, they showed that the wetting layer at very early times exhibited power-law growth in the potential-dependent regime. The growth exponent was found to be strongly dependent on $ n $. It then crosses over to a diffusive regime. Note that the above-mentioned studies were based on the \emph{diffusion-driven} model. However, for fluid mixtures, the hydrodynamics effects become relevant, and faster growth of the wetting layer is observed at late times \cite{tanaka01,SSJ01,puri05}. Jaiswal \emph{et al.} \cite{PSS12} carried out molecular dynamics simulation to study the kinetics of SDSD. They deduced the growth laws for the wetting-layer thickness $R_1$ $\sim$ $t^\alpha$, with $\alpha \approx 1/3$ (diffusive regime) and $\alpha \approx 1$ (viscous hydrodynamic regime). The usual phase separation was observed for bulk domains. 

\par In the present work, we perform comprehensive molecular dynamics (MD) simulation to investigate the effects of long-ranged power-law and short-ranged exponential surface potentials on the early-stage dynamics of SDSD. We primarily focus on the potential-dependent growth of the wetting layer for symmetric (critical) binary mixtures and its crossover to the diffusive regime via an intermediate \emph{fast-mode kinetics}. The fast mode was observed experimentally for polymer blends as well as low-molecular-weight critical mixtures by Wiltzius, Cumming, and coworkers \cite{PA91, APF92,BCA93}. For the long-ranged interactions, we observe that the growth laws for $ R_1(t) $ exhibit power-law behavior in various regimes. As the growth of the wetting layer unfolds, distinct and noteworthy crossovers are observed in the following sequence: \emph{potential-dependent regime $ \rightarrow $  fast-mode regime  $ \rightarrow $ diffusive regime  $ \rightarrow $ hydrodynamic regime}. In contrast, for the case with short-ranged wall potential, a logarithmic growth of the wetting layer is observed that again crosses over to a fast mode and subsequently to the well-established diffusive and hydrodynamic regimes.

%\section{\label{sec:md}Methodology}
\emph{Methodology.} To study the surface-directed spinodal decomposition (SDSD), we carry out molecular dynamics (MD) simulations using LAMMPS \cite{plimpton95} software package.  %which efficiently parallelizes the computationally extensive workload to adequate compute cores and thus, considerably reduces the simulation time. 
We consider a binary mixture of $ A $ and $ B $ type particles in a box of volume $V = L\times L \times D$ with an equal number of $ A $ and $ B $ ($N_A=N_B=N/2$) particles. We apply periodic boundary conditions in $ x, y $-directions, and a wall is placed in the $ z $-direction at $z=0$. The wall is impenetrable to either of the particles, and the simulation box constitutes a semi-infinite geometry. 

\par The particle-wall interaction is modeled via an integrated Lennard-Jones (LJ) potential ($u_w$), given as
\begin{equation}\label{eq2:wall}
u_w(z)=\frac{2\pi \rho_N \sigma^3}{3}\left[ \frac{2\epsilon_r}{15}\left( \frac{\sigma}{z^\prime}\right)^9 + \delta_\alpha V(z^\prime)\right], \;\;\;\;\; z^\prime = z + \sigma/2
\end{equation}
with $V(z^\prime)$ representing the attractive potential and has been employed in two forms in the present study. These forms mimic the long-ranged and short-ranged surface interactions with the particles:
\begin{equation}\label{eq3:att}
V(z^\prime) =
\begin{dcases}
V_{long} = -\epsilon_a\left(\dfrac{\sigma}{z^\prime}\right)^n , & \\
V_{short} = -\epsilon_a \exp\left(-z^\prime/z_0\right). 
\end{dcases}
\end{equation}
The parameters $\epsilon_r$ and $\epsilon_a$ set the energy scales for the wall-particle repulsion and attraction, respectively. $\sigma$ is the LJ diameter, and $\rho_N$ is the fluid density ($\rho_N=N/L^2D$). The attractive term in the potential in Eq.~\eqref{eq2:wall} is switched on or off using $\delta_\alpha$ for ``$\alpha$" type of particles. Therefore, to have an $ A$-rich wetting layer at $ z=0 $, we set $\delta_A=1$ and $\delta_B=0$. Moreover, the variable $ z^\prime $ in Eq.~\eqref{eq2:wall} is defined as $z^\prime = z + \sigma/2$ so that the singularity in $ u_w(z) $ occurs outside the simulation box at $z=-\sigma/2$. We place a confining surface at $z=D$ with $z^\prime=D+\sigma/2-z$. However, in this case, we set $\delta_A=\delta_B=0$  so that both $ A $ and $ B $ particles are repelled. The particle-particle interaction and the other simulation details, including thermostat, integrator, etc., are provided in the Supplemental Material, Sec. I \cite{sm}.

\begin{figure}[!ht]
	\centering
	\includegraphics[width=0.45\textwidth]{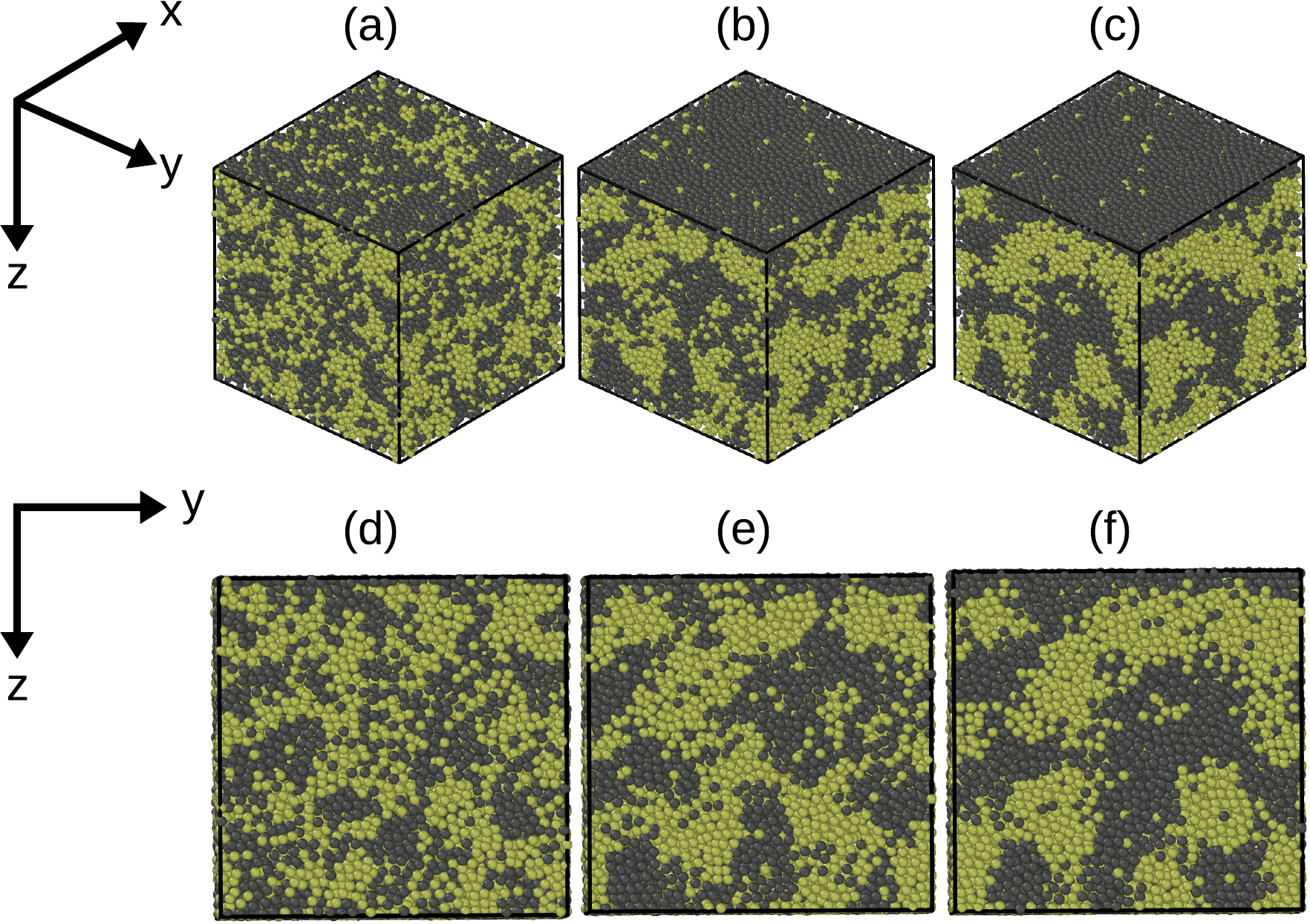}
	\caption{Upper panel: Evolution snapshots for a binary Lennard-Jones mixture ($ A+B $) undergoing surface-directed spinodal decomposition (SDSD) within a box of dimensions $L^2 \times D $, with $L=D=32$. These snapshots correspond to $t=40$ (a), $t=200$ (b), and $t=400$ (c). An impenetrable surface at $z=0$ attracts $ A $ particles (marked in gray). The choice of interaction coefficients between the wall and particles, in Eq.~\eqref{eq3:att}, produces completely wet (CW) morphology at equilibrium. The system is quenched from a high temperature (corresponding to a homogeneous initial state) to $T=1.0 < T_c$. The other simulation details are mentioned in the text and \cite{sm}. Lower panel: $yz$ cross-sections of the  snapshots shown in the upper panel at $x=L$.}
	\label{fig1}
\end{figure}

\emph{Results and Discussion.} First, we present results from our molecular dynamics (MD) simulations for a long-ranged power-law potential defined in Eq.~\eqref{eq3:att}. The interaction potential parameters are $\epsilon_r=0.6$ and $\epsilon_a=0.5$ corresponding to a completely wet (CW) morphology in equilibrium \cite{SK10}. The simulation box is $L^2 \times D$ with $L=D=32$, i.e., the total number of particles is $N=32768$. We present the evolution snapshots of the system in Fig.~\ref{fig1} for the power-law wall potential with an exponent $n=3$. Noticeably, the bulk segregates into $ A $-rich and $ B $-rich phases with bicontinuous domain morphology. Moreover, the surface at $z=0$ starts populating with $ A $ particles, whereas the $ B $ particles migrate towards the bulk. This leads to the formation of an $ A $-rich layer at the surface with a $B$-rich depletion layer on top of it. 

\par Further, the bulk domains as well as both the wetting and depletion layers grow with time (for instance, in Fig.~\ref{fig1}(c) and (f) corresponding to $ t=400 $,  a thicker $ A $-rich wetting layer is seen at the surface in comparison to Fig.~\ref{fig1}(a) and (d) which correspond to $ t=40 $). Remarkably, the transport of $A$-rich bulk tubes into the wetting layer is evident in Fig.~\ref{fig1}(f). This flush of wettable tubes spreads along the surface and leads to the radial thickening of one end of the tube connected to the wetting layer (see Fig.~S1 and movie~SM1 \cite{sm}). The resultant wetting-layer growth is much faster than the diffusive dynamics and is attributed to the hydrodynamic effects \cite{tanaka93-2,tanaka93,tanaka93-3,tanaka01}. We end our discussion on Fig.~\ref{fig1} by highlighting a similar observation (though, at a much earlier time) of hydrodynamic pumping of bulk tubes into the wetting layer (see Fig.~S2 and movie~SM2 \cite{sm}). This is the first numerical observation of what has been experimentally referred to as \emph{fast-mode} kinetics \cite{PA91,APF92}. There has been some controversy regarding the existence of this regime of wetting kinetics \cite{tanaka93,PWA94prl,PWA93,troian94}. Our MD results in this paper demonstrate the presence of a universal fast mode and it is the central result of this Letter.
 %%%%%%%%%%%%%%%%%%%%%%%%%%%%%%%%%%%%%%%%%%%%
 %%%%%%%%%%%%%%%%%%%%%%%%%%%%%%%%%%%%%%%%%%%%
 \par
\begin{figure}[!ht]
 	\centering
 	\includegraphics[width=0.45\textwidth]{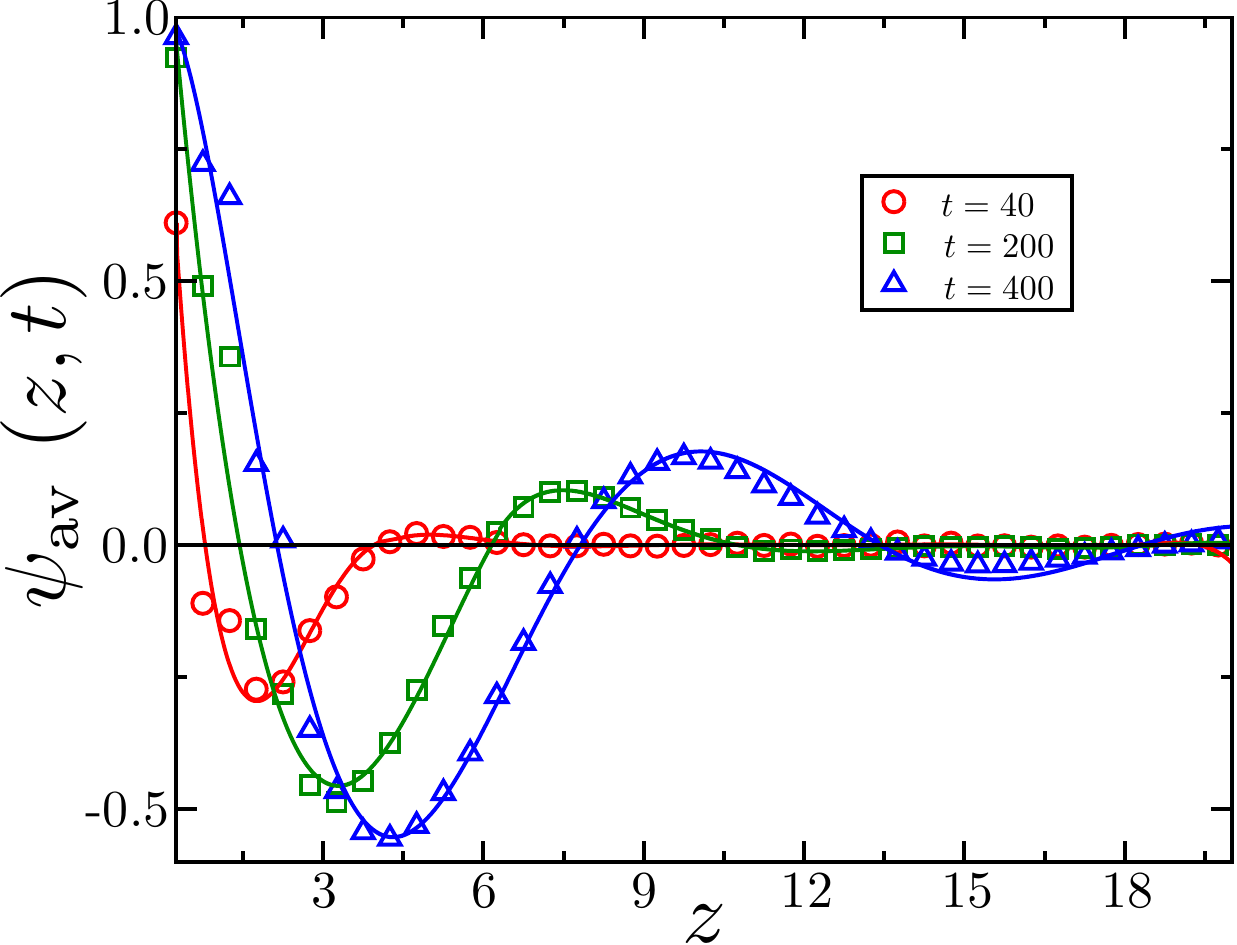}
 	\caption{Laterally-averaged order parameter profiles $\psi_{\text{av}} (z,t)$ vs. $z$ for the evolution shown in Fig.~\ref{fig1} at $t = 40, 200$, and $400$. The solid lines through the data set are guides to the eye. }
 	\label{fig2}
 \end{figure}

 The multilayered morphology shown in Fig.~\ref{fig1}  results in the emergence of a wave-like structure known as the SDSD wave and can be corroborated from the depth profiles presented in Fig.~\ref{fig2}. We employ the laterally-averaged order parameter for depth profiling of the morphology formed at the surface. The order parameter, $\psi$,  is calculated from the local densities $n_A$ and $n_B$ as $ \psi(x,y,z,t) = (n_A - n_B) / (n_A + n_B)$ and is then averaged over the $xy$ plane in a fixed layer thickness $\Delta z=0.5$ centered at different $z$ values to produce the laterally-averaged order parameter $ \psi_{\text{av}}(z,t) $. We finally average these profiles over $400$ independent runs to improve the statistics. We plot $ \psi_{\text{av}}(z,t) $ vs. $ z $ in Fig.~\ref{fig2} for three different times $ t=40, 200, $  and $400$ corresponding to potential-dependent, diffusive, and viscous hydrodynamic regimes, respectively as will be discussed shortly. The SDSD wave originates at the surface (where $ \psi_{\text{av}}(z,t)\approx 1 $) and vanishes into the bulk (where $ \psi_{\text{av}}(z,t)\approx 0 $). Such depth profiles exhibit a structured oscillatory behavior at the surface and offer an experimental counterpart of depth-profiling techniques \cite{krausch95}.
 %%%%%%%%%%%%%%%%%%%%%%%%%%%%%%%%%%%%%%%%%%%%
 %%%%%%%%%%%%%%%%%%%%%%%%%%%%%%%%%%%%%%%%%%%%

\par
\begin{figure}[h]
	\centering
	\includegraphics[width=0.45\textwidth]{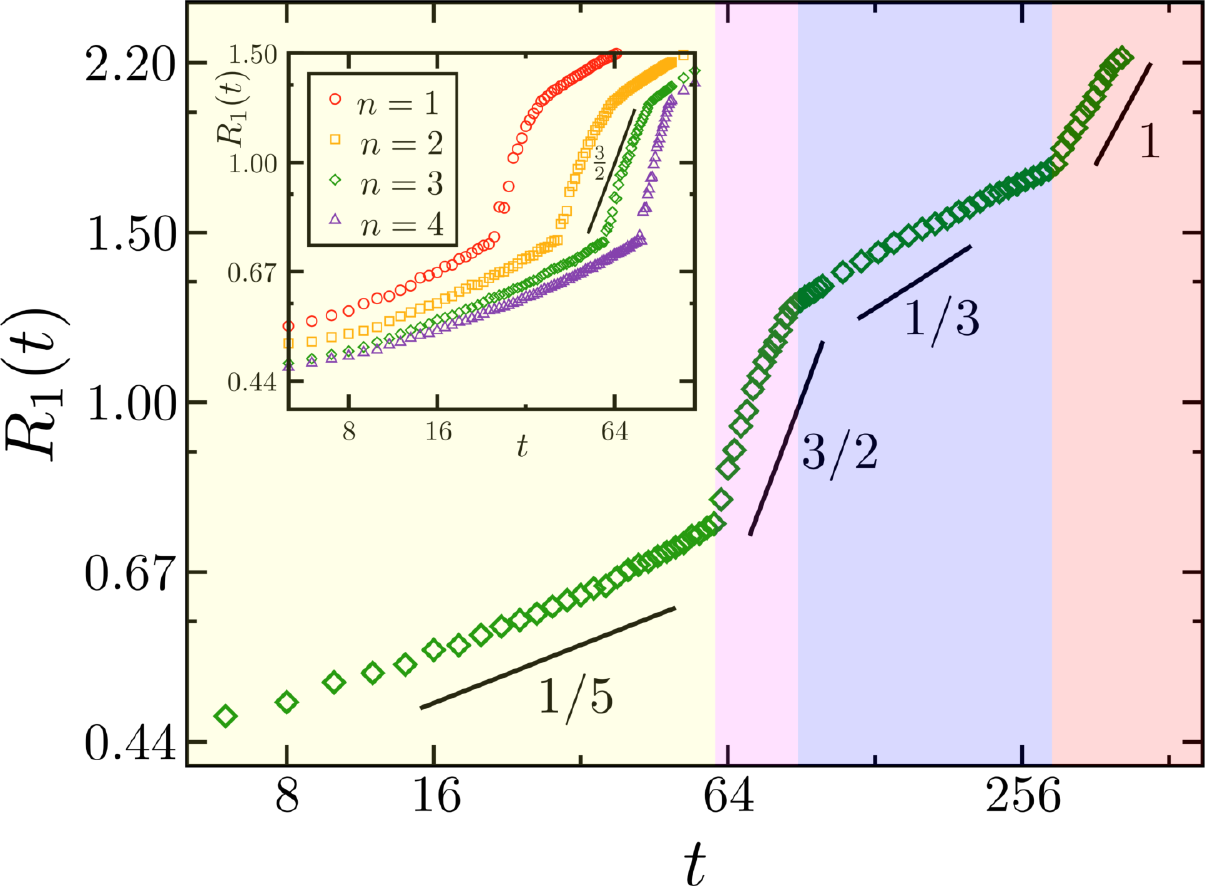}
	\caption{The time dependence of $R_1(t)$ on a log-log scale for the CW morphology. The results are presented for the power-law surface potential, $-\epsilon_a/z^n$. Main figure: the wetting layer growth when the surface potential is specified by $n=3$, corresponding to van der Waals' interaction. The straight lines having slopes $1/5$, $1/3$, and $1$ correspond to the potential-dependent, diffusive, and hydrodynamic regimes, respectively. Inset: $R_1(t)$ vs. $t$ for the potential specified by $n=1,2,3$, and $4$. The solid line with slope $3/2$ suggests the fast mode observed by Wiltzius and Cumming \cite{PA91}. }
	\label{fig3}
\end{figure}

 We examine the evolution of the SDSD profile by computing the first zero-crossing, $R_1(t)$, from the laterally-averaged order parameter shown in Fig.~\ref{fig2}. The quantity $R_1(t)$ characterizes the wetting-layer thickness, and its evolution with time $t$ is shown in Fig.~\ref{fig3}. We observe four distinct time regimes shaded in different colors: potential dependent, fast-mode, diffusive, and viscous hydrodynamic regimes as time progresses. Each regime exhibits a power-law growth $R_1(t) \sim t^\alpha$ (where $\alpha$ denotes the growth exponent) and is punctuated by a sharp crossover. In our previous work on MD of SDSD \cite{PSS12-epl}, we established such a crossover from the universal diffusive regime ($\alpha = 1/3$) to the viscous hydrodynamic regime ($\alpha = 1$). However, we did not explore the early-time behavior of SDSD characterized by the potential-dependent regime and the fast mode. 

\par The primary objectives of this work are to obtain the growth laws for the early-time kinetics. The growth exponent for the wetting-layer thickness at early times is $\alpha \approx 1/5$ for the power-law surface potential with $n = 3$. This is consistent with the phenomenological theory of SDSD \cite{SK92} discussed in the Supplemental Material, Sec. II \cite{sm}.  As time evolves, the potential-dependent growth crosses over to a very rapid regime with an exponent $\alpha \approx 3/2$. The short-lived accelerated growth in $ t \in (60,100)$ (see Fig.~\ref{fig3}) is the first numerical observation of the \emph{fast-mode kinetics} observed in experiments by Wiltzius, Cumming, and others \cite{PA91, APF92,BCA93}. 
 
%%%%%%%%%%%%%%%%%%%%%%%%%%%%%%%%%%%%%%%%%%%% 
%%%%%%%%%%%%%%%%%%%%%%%%%%%%%%%%%%%%%%%%%%%%

\par How do we comprehend the fast-mode kinetics seen for the growth in the wetting-layer thickness? The earlier studies reported in the literature were not conclusive: the primary reason was that the regime was very short-lived and was suggested to be transient \cite{PA91,tanaka01,HT00}. To this end, we perform our simulations for different values of $n$ (as defined in $V(z) \sim z^{-n}$), and also examine the effect of varying the range of the surface potential. We show these results in the inset of Fig.~\ref{fig3} where we plot $R_1(t)$ vs. $t$ for $n=1,2,3,$ and $4$ ($n=\infty$ will correspond to a $\delta$-function potential). Conspicuously, the crossover from the potential-dependent regime to the universal fast-mode regime is seen for all the values of $n$. However, the regime is indeed short-lived, following the experiments \cite{PA91,tanaka01,HT00}. It is worthwhile to mention here that the growth exponent for $R_1(t)$ is nearly the same (i.e., $R_1(t) \sim t^\alpha$ with $\alpha \approx 3/2$) for all $n$ values in the fast-mode regime. Therefore, we conclude that the fast-mode kinetics also follows a universal power-law growth with an exponent $3/2$, in addition to the diffusive $(\alpha \approx 1/3)$ and viscous hydrodynamic $(\alpha \approx 1)$ regimes. In contrast, we find different power-law behaviors for early time (before the fast mode triggers) as we change $n$. Consequently, we label this as the \emph{potential-dependent regime} and will investigate it in detail shortly. 

Further, as $n$ increases, the decrease in the range of $V(z)$ results in a reduced surface effect on the wetting-layer growth. This increases the crossover time $t_c$ (see inset of Fig.~\ref{fig3}) to the fast mode. Nevertheless, the values of $R_1(t) \sim 0.75 \,\sigma$ at the onset of fast mode and $R_1(t) \sim 1.25 \,\sigma$ at the onset of diffusion-driven growth do not change with varying $n$ and remain almost fixed. We  further make an important observation for the accelerated growth of $R_1(t) \sim 0.75\,\sigma$ to $R_1(t) \sim 1.25\,\sigma$ implying the completion of first wetting layer structure with an average thickness $\sim \sigma$. This completion of the first wetting layer (\emph{coating dynamics}) is assisted by the domains near the surface via the hydrodynamic pumping mechanism (see Fig.~S2 and movie~SM2 \cite{sm}) \cite{tanaka93}. 

Before addressing the subsequent growth regime, we recall Tanaka's argument \cite{tanaka93, tanaka01, tanaka93-3} for the observed large exponent $\alpha \approx 3/2$ to complete our discourse on the faster growth regime. Due to the pressure gradient between the interconnected bicontinuous tubes of the wetting component and the surface, a flux estimated as $J \sim (\gamma / \eta) L^2$ is developed. Here, $\gamma, \eta$, and $L$ denote surface tension, viscosity, and tube size, respectively. In the case of strong wettability, the tubes are hydrodynamically flushed towards the surface, and the coating dynamic is observed via lateral spreading of a droplet near the surface. The size of the droplet $L_s(t)$ in 3D grows as $dL_s^2/dt \sim (\gamma / \eta) L^2$. Now, applying Siggia's law, $L(t) \sim (\gamma / \eta) t$, the growth of the wetting-layer thickness is obtained as $R_1(t) \sim L_s(t) \sim [(\gamma / \eta) \,t]^{3/2}$. This agrees with the earlier experimental observations \cite{PA91, APF92,BCA93} and our present MD simulations.

\par
\begin{figure}[!ht]
	\centering
\includegraphics[width=0.45\textwidth]{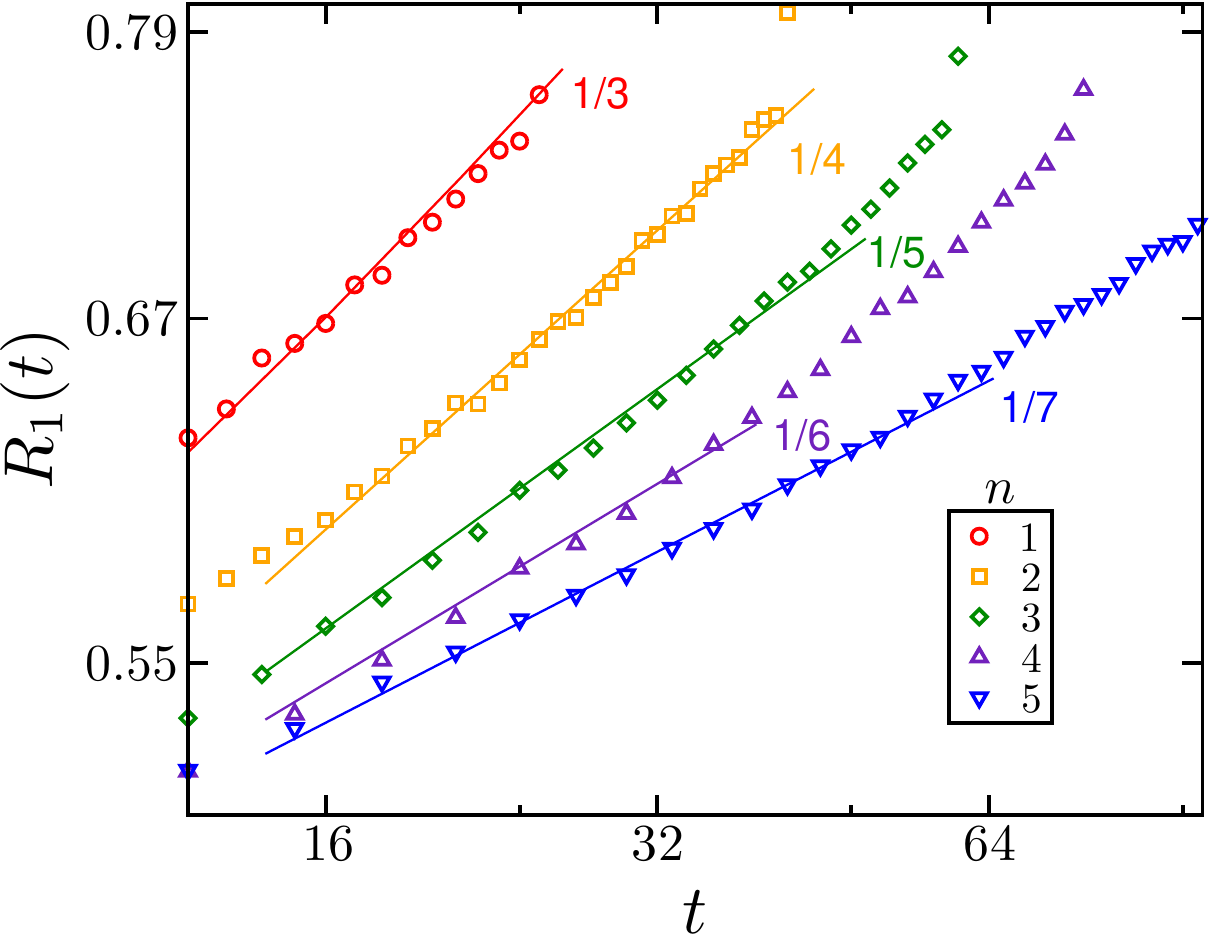}
	\caption{Growth of $R_1(t)$ with $t$ on a log-log scale for different $n$ defined in the power-law surface potential, $-\epsilon_a/z^n$. The results are shown for $t<100$, illustrating the potential-dependent regime. The colored lines have slopes $1/3$, $1/4$, $1/5$, $1/6$, and $1/7$ for $n=1$, $2$, $3$, $4$, and $5$, respectively, as shown in the legend.}
	\label{fig4}
\end{figure}

 Next, we turn our attention to the potential-dependent growth regime. In Fig.~\ref{fig4}, we plot $R_1(t)$ vs. $t$, restricting the measurement to very early times. We investigate the growth kinetics for different values of the exponent $n$ (as defined in $V(z) \sim z^{-n}$). It is evident that $R_1(t) \sim t^\alpha$, where the growth exponent $\alpha$ depends on the power $n$ via $\alpha = 1/(n+2)$, {\it{i.e.}}, $R_1(t) \sim t^{1/(n+2)}$. Our simulation results for five different values of $n$ are in excellent agreement with the predictions of the phenomenological theory on SDSD \cite{SK01, puri05} summarized in the Supplemental Material, Sec. II \cite{sm}. 
 %%%%%%%%%%%%%%%%%%%%%%%%%%%%%%%%%%%%%%%%%%%%
 %%%%%%%%%%%%%%%%%%%%%%%%%%%%%%%%%%%%%%%%%%%%

To further test the universality of the fast-mode kinetics, we simulate the present system with an exponential wall potential. Here, in contrast, the wetting-layer growth exhibits a logarithmic behavior at very early times, as observed earlier \cite{puri05}. Nevertheless, it crosses over to the universal fast-mode regime. Further details, along with the numerical results, are provided in the Supplemental Material,  Sec. V \cite{sm}.

\emph{Final remarks.} In conclusion, we have performed extensive molecular dynamics (MD) simulations of surface-directed spinodal decomposition (SDSD) of symmetric fluid mixtures at a wetting surface. The symmetry-breaking surface potentials studied here are long-ranged (power-law) and short-ranged (exponential) potentials. We present results for various growth regimes of wetting-layer thickness $R_1(t)$ extracted from the SDSD depth profiles. Each segment of the time dependence of $R_1(t)$ reflects a different transport mechanism inaction responsible for different growth laws. The prime focus of this work is to settle any controversy regarding the existence of a universal fast-mode. In our simulations, this is unambiguously seen as a crossover regime between the potential-dependent growth and the universal diffusive dynamics. Our evolution snapshots and movies show these fast mode results from the rapid growth of droplets preceding the formation of the wetting layer. This is enabled by the emergence of bulk fluid tubes, which rapidly drain material into the surface droplets. The corresponding growth exponent is $\approx 3/2$. This is consistent with the experimental works on binary polymer blends, a fluid mixture of guaiacol and glycerol-water, etc., by Wiltzius, Cumming, and others \cite{PA91, APF92,BCA93} and theoretical predictions by Tanaka \cite{tanaka93-2, tanaka93,tanaka93-3}. Regardless of the surface-field form, its range, and strength, we find a fast mode with the universal growth exponent $\approx 3/2$. For the long-ranged surface potential, $V(z)\sim z^{-n}$, $R_1(t)$ displays a power-law dependence with $R_1(t) \sim t^{1/(n+2)}$ at very early times, supporting the phenomenological theory \cite{SK01}. Finally, in the case of short-ranged potential, $R_1(t)$ shows a logarithmic growth followed by fast-mode, diffusive, and hydrodynamic regime. To the best of our understanding, the results (for both short-ranged and long-ranged interactions) presented in this work are the first atomistic simulations that access all the growth regimes of wetting kinetics.
%\acknowledgments

\par \emph{Acknowledgments.} M.P. acknowledges funding from the Department of Science and Technology (DST), India, through a SERB-ECR Grant No. \emph{ECR/2017/003091}.

\par All authors declare they have no competing interests.

\bibliographystyle{apsrev4-2}
%\bibliography{fastmode.bbl}

\begin{thebibliography}{86}%
	\makeatletter
	\providecommand \@ifxundefined [1]{%
		\@ifx{#1\undefined}
	}%
	\providecommand \@ifnum [1]{%
		\ifnum #1\expandafter \@firstoftwo
		\else \expandafter \@secondoftwo
		\fi
	}%
	\providecommand \@ifx [1]{%
		\ifx #1\expandafter \@firstoftwo
		\else \expandafter \@secondoftwo
		\fi
	}%
	\providecommand \natexlab [1]{#1}%
	\providecommand \enquote  [1]{``#1''}%
	\providecommand \bibnamefont  [1]{#1}%
	\providecommand \bibfnamefont [1]{#1}%
	\providecommand \citenamefont [1]{#1}%
	\providecommand \href@noop [0]{\@secondoftwo}%
	\providecommand \href [0]{\begingroup \@sanitize@url \@href}%
	\providecommand \@href[1]{\@@startlink{#1}\@@href}%
	\providecommand \@@href[1]{\endgroup#1\@@endlink}%
	\providecommand \@sanitize@url [0]{\catcode `\\12\catcode `\$12\catcode
		`\&12\catcode `\#12\catcode `\^12\catcode `\_12\catcode `\%12\relax}%
	\providecommand \@@startlink[1]{}%
	\providecommand \@@endlink[0]{}%
	\providecommand \url  [0]{\begingroup\@sanitize@url \@url }%
	\providecommand \@url [1]{\endgroup\@href {#1}{\urlprefix }}%
	\providecommand \urlprefix  [0]{URL }%
	\providecommand \Eprint [0]{\href }%
	\providecommand \doibase [0]{https://doi.org/}%
	\providecommand \selectlanguage [0]{\@gobble}%
	\providecommand \bibinfo  [0]{\@secondoftwo}%
	\providecommand \bibfield  [0]{\@secondoftwo}%
	\providecommand \translation [1]{[#1]}%
	\providecommand \BibitemOpen [0]{}%
	\providecommand \bibitemStop [0]{}%
	\providecommand \bibitemNoStop [0]{.\EOS\space}%
	\providecommand \EOS [0]{\spacefactor3000\relax}%
	\providecommand \BibitemShut  [1]{\csname bibitem#1\endcsname}%
	\let\auto@bib@innerbib\@empty
	%</preamble>
	\bibitem [{\citenamefont {Puri}\ and\ \citenamefont {Wadhawan}(2009)}]{VS09}%
	\BibitemOpen
	\bibfield  {author} {\bibinfo {author} {\bibfnamefont {S.}~\bibnamefont
			{Puri}}\ and\ \bibinfo {author} {\bibfnamefont {V.}~\bibnamefont
			{Wadhawan}},\ }\href {https://doi.org/10.1201/9781420008364} {\emph {\bibinfo
			{title} {Kinetics of Phase Transitions}}},\ \bibinfo {edition} {1st}\ ed.\
	(\bibinfo  {publisher} {CRC Press},\ \bibinfo {address} {Boca Raton},\
	\bibinfo {year} {2009})\BibitemShut {NoStop}%
	\bibitem [{\citenamefont {Bray}(2002)}]{bray02}%
	\BibitemOpen
	\bibfield  {author} {\bibinfo {author} {\bibfnamefont {A.~J.}\ \bibnamefont
			{Bray}},\ }\href {https://doi.org/10.1080/00018730110117433} {\bibfield
		{journal} {\bibinfo  {journal} {Advances in Physics}\ }\textbf {\bibinfo
			{volume} {51}},\ \bibinfo {pages} {481} (\bibinfo {year} {2002})}\BibitemShut
	{NoStop}%
	\bibitem [{\citenamefont {Onuki}(2002)}]{onuki02}%
	\BibitemOpen
	\bibfield  {author} {\bibinfo {author} {\bibfnamefont {A.}~\bibnamefont
			{Onuki}},\ }\href {https://doi.org/10.1017/CBO9780511534874} {\emph {\bibinfo
			{title} {Phase Transition Dynamics}}}\ (\bibinfo  {publisher} {Cambridge
		University Press},\ \bibinfo {year} {2002})\BibitemShut {NoStop}%
	\bibitem [{\citenamefont {Geoghegan}\ \emph {et~al.}(1995)\citenamefont
		{Geoghegan}, \citenamefont {Jones},\ and\ \citenamefont {Clough}}]{MRA95}%
	\BibitemOpen
	\bibfield  {author} {\bibinfo {author} {\bibfnamefont {M.}~\bibnamefont
			{Geoghegan}}, \bibinfo {author} {\bibfnamefont {R.~A.~L.}\ \bibnamefont
			{Jones}},\ and\ \bibinfo {author} {\bibfnamefont {A.~S.}\ \bibnamefont
			{Clough}},\ }\href {https://doi.org/10.1063/1.470506} {\bibfield  {journal}
		{\bibinfo  {journal} {The Journal of Chemical Physics}\ }\textbf {\bibinfo
			{volume} {103}},\ \bibinfo {pages} {2719} (\bibinfo {year}
		{1995})}\BibitemShut {NoStop}%
	\bibitem [{\citenamefont {Jones}\ \emph {et~al.}(1991)\citenamefont {Jones},
		\citenamefont {Norton}, \citenamefont {Kramer}, \citenamefont {Bates},\ and\
		\citenamefont {Wiltzius}}]{RLE91}%
	\BibitemOpen
	\bibfield  {author} {\bibinfo {author} {\bibfnamefont {R.~A.~L.}\
			\bibnamefont {Jones}}, \bibinfo {author} {\bibfnamefont {L.~J.}\ \bibnamefont
			{Norton}}, \bibinfo {author} {\bibfnamefont {E.~J.}\ \bibnamefont {Kramer}},
		\bibinfo {author} {\bibfnamefont {F.~S.}\ \bibnamefont {Bates}},\ and\
		\bibinfo {author} {\bibfnamefont {P.}~\bibnamefont {Wiltzius}},\ }\href
	{https://doi.org/10.1103/PhysRevLett.66.1326} {\bibfield  {journal} {\bibinfo
			{journal} {Phys. Rev. Lett.}\ }\textbf {\bibinfo {volume} {66}},\ \bibinfo
		{pages} {1326} (\bibinfo {year} {1991})}\BibitemShut {NoStop}%
	\bibitem [{\citenamefont {Haase}\ and\ \citenamefont {Brujic}(2014)}]{MJ14}%
	\BibitemOpen
	\bibfield  {author} {\bibinfo {author} {\bibfnamefont {M.~F.}\ \bibnamefont
			{Haase}}\ and\ \bibinfo {author} {\bibfnamefont {J.}~\bibnamefont {Brujic}},\
	}\href {https://doi.org/10.1002/anie.201406040} {\bibfield  {journal}
		{\bibinfo  {journal} {Angewandte Chemie International Edition}\ }\textbf
		{\bibinfo {volume} {53}},\ \bibinfo {pages} {11793} (\bibinfo {year}
		{2014})}\BibitemShut {NoStop}%
	\bibitem [{\citenamefont {Kim}\ and\ \citenamefont {Weitz}(2011)}]{SD11}%
	\BibitemOpen
	\bibfield  {author} {\bibinfo {author} {\bibfnamefont {S.-H.}\ \bibnamefont
			{Kim}}\ and\ \bibinfo {author} {\bibfnamefont {D.~A.}\ \bibnamefont
			{Weitz}},\ }\href {https://doi.org/10.1002/anie.201102946} {\bibfield
		{journal} {\bibinfo  {journal} {Angewandte Chemie International Edition}\
		}\textbf {\bibinfo {volume} {50}},\ \bibinfo {pages} {8731} (\bibinfo {year}
		{2011})}\BibitemShut {NoStop}%
	\bibitem [{\citenamefont {Bouttes}\ \emph {et~al.}(2014)\citenamefont
		{Bouttes}, \citenamefont {Gouillart}, \citenamefont {Boller}, \citenamefont
		{Dalmas},\ and\ \citenamefont {Vandembroucq}}]{DEE14}%
	\BibitemOpen
	\bibfield  {author} {\bibinfo {author} {\bibfnamefont {D.}~\bibnamefont
			{Bouttes}}, \bibinfo {author} {\bibfnamefont {E.}~\bibnamefont {Gouillart}},
		\bibinfo {author} {\bibfnamefont {E.}~\bibnamefont {Boller}}, \bibinfo
		{author} {\bibfnamefont {D.}~\bibnamefont {Dalmas}},\ and\ \bibinfo {author}
		{\bibfnamefont {D.}~\bibnamefont {Vandembroucq}},\ }\href
	{https://doi.org/10.1103/PhysRevLett.112.245701} {\bibfield  {journal}
		{\bibinfo  {journal} {Phys. Rev. Lett.}\ }\textbf {\bibinfo {volume} {112}},\
		\bibinfo {pages} {245701} (\bibinfo {year} {2014})}\BibitemShut {NoStop}%
	\bibitem [{\citenamefont {Tateno}\ and\ \citenamefont {Tanaka}(2021)}]{MH21}%
	\BibitemOpen
	\bibfield  {author} {\bibinfo {author} {\bibfnamefont {M.}~\bibnamefont
			{Tateno}}\ and\ \bibinfo {author} {\bibfnamefont {H.}~\bibnamefont
			{Tanaka}},\ }\href {https://doi.org/10.1038/s41467-020-20734-8} {\bibfield
		{journal} {\bibinfo  {journal} {Nature Communications}\ }\textbf {\bibinfo
			{volume} {12}},\ \bibinfo {pages} {912} (\bibinfo {year} {2021})}\BibitemShut
	{NoStop}%
	\bibitem [{\citenamefont {Lu}\ \emph {et~al.}(2009)\citenamefont {Lu},
		\citenamefont {Yan},\ and\ \citenamefont {Texter}}]{JFJ09}%
	\BibitemOpen
	\bibfield  {author} {\bibinfo {author} {\bibfnamefont {J.}~\bibnamefont
			{Lu}}, \bibinfo {author} {\bibfnamefont {F.}~\bibnamefont {Yan}},\ and\
		\bibinfo {author} {\bibfnamefont {J.}~\bibnamefont {Texter}},\ }\href
	{https://doi.org/10.1016/j.progpolymsci.2008.12.001} {\bibfield  {journal}
		{\bibinfo  {journal} {Progress in Polymer Science}\ }\textbf {\bibinfo
			{volume} {34}},\ \bibinfo {pages} {431} (\bibinfo {year} {2009})}\BibitemShut
	{NoStop}%
	\bibitem [{\citenamefont {Koster}\ \emph {et~al.}(2005)\citenamefont {Koster},
		\citenamefont {Smits}, \citenamefont {Mihailetchi},\ and\ \citenamefont
		{Blom}}]{LEV05}%
	\BibitemOpen
	\bibfield  {author} {\bibinfo {author} {\bibfnamefont {L.~J.~A.}\
			\bibnamefont {Koster}}, \bibinfo {author} {\bibfnamefont {E.~C.~P.}\
			\bibnamefont {Smits}}, \bibinfo {author} {\bibfnamefont {V.~D.}\ \bibnamefont
			{Mihailetchi}},\ and\ \bibinfo {author} {\bibfnamefont {P.~W.~M.}\
			\bibnamefont {Blom}},\ }\href {https://doi.org/10.1103/PhysRevB.72.085205}
	{\bibfield  {journal} {\bibinfo  {journal} {Phys. Rev. B}\ }\textbf {\bibinfo
			{volume} {72}},\ \bibinfo {pages} {085205} (\bibinfo {year}
		{2005})}\BibitemShut {NoStop}%
	\bibitem [{\citenamefont {Liu}\ \emph {et~al.}(2021)\citenamefont {Liu},
		\citenamefont {Cao}, \citenamefont {Zhu}, \citenamefont {Wang}, \citenamefont
		{Tang}, \citenamefont {Wu}, \citenamefont {Zou}, \citenamefont {Feng},\ and\
		\citenamefont {Ma}}]{NYY21}%
	\BibitemOpen
	\bibfield  {author} {\bibinfo {author} {\bibfnamefont {N.}~\bibnamefont
			{Liu}}, \bibinfo {author} {\bibfnamefont {Y.}~\bibnamefont {Cao}}, \bibinfo
		{author} {\bibfnamefont {Y.-L.}\ \bibnamefont {Zhu}}, \bibinfo {author}
		{\bibfnamefont {Y.-J.}\ \bibnamefont {Wang}}, \bibinfo {author}
		{\bibfnamefont {Y.-L.}\ \bibnamefont {Tang}}, \bibinfo {author}
		{\bibfnamefont {B.}~\bibnamefont {Wu}}, \bibinfo {author} {\bibfnamefont
			{M.-J.}\ \bibnamefont {Zou}}, \bibinfo {author} {\bibfnamefont {Y.-P.}\
			\bibnamefont {Feng}},\ and\ \bibinfo {author} {\bibfnamefont {X.-L.}\
			\bibnamefont {Ma}},\ }\href {https://doi.org/10.1021/acsami.1c06649}
	{\bibfield  {journal} {\bibinfo  {journal} {ACS Applied Materials \&
				Interfaces}\ }\textbf {\bibinfo {volume} {13}},\ \bibinfo {pages} {31001}
		(\bibinfo {year} {2021})}\BibitemShut {NoStop}%
	\bibitem [{\citenamefont {Xiang}\ \emph {et~al.}(2021)\citenamefont {Xiang},
		\citenamefont {Zhang}, \citenamefont {Xin}, \citenamefont {An}, \citenamefont
		{Niu}, \citenamefont {Mardani}, \citenamefont {Siegrist}, \citenamefont {Lu},
		\citenamefont {Goddard}, \citenamefont {Man}, \citenamefont {Wang},\ and\
		\citenamefont {Han}}]{ZLY21}%
	\BibitemOpen
	\bibfield  {author} {\bibinfo {author} {\bibfnamefont {Z.}~\bibnamefont
			{Xiang}}, \bibinfo {author} {\bibfnamefont {L.}~\bibnamefont {Zhang}},
		\bibinfo {author} {\bibfnamefont {Y.}~\bibnamefont {Xin}}, \bibinfo {author}
		{\bibfnamefont {B.}~\bibnamefont {An}}, \bibinfo {author} {\bibfnamefont
			{R.}~\bibnamefont {Niu}}, \bibinfo {author} {\bibfnamefont {M.}~\bibnamefont
			{Mardani}}, \bibinfo {author} {\bibfnamefont {T.}~\bibnamefont {Siegrist}},
		\bibinfo {author} {\bibfnamefont {J.}~\bibnamefont {Lu}}, \bibinfo {author}
		{\bibfnamefont {R.~E.}\ \bibnamefont {Goddard}}, \bibinfo {author}
		{\bibfnamefont {T.}~\bibnamefont {Man}}, \bibinfo {author} {\bibfnamefont
			{E.}~\bibnamefont {Wang}},\ and\ \bibinfo {author} {\bibfnamefont
			{K.}~\bibnamefont {Han}},\ }\href
	{https://doi.org/10.1016/j.matdes.2020.109383} {\bibfield  {journal}
		{\bibinfo  {journal} {Materials \& Design}\ }\textbf {\bibinfo {volume}
			{199}},\ \bibinfo {pages} {109383} (\bibinfo {year} {2021})}\BibitemShut
	{NoStop}%
	\bibitem [{\citenamefont {Deguchi}\ \emph {et~al.}(2020)\citenamefont
		{Deguchi}, \citenamefont {Nakamura},\ and\ \citenamefont {Ohno}}]{YNH20}%
	\BibitemOpen
	\bibfield  {author} {\bibinfo {author} {\bibfnamefont {Y.}~\bibnamefont
			{Deguchi}}, \bibinfo {author} {\bibfnamefont {N.}~\bibnamefont {Nakamura}},\
		and\ \bibinfo {author} {\bibfnamefont {H.}~\bibnamefont {Ohno}},\ }\href
	{https://doi.org/10.1016/j.seppur.2020.117286} {\bibfield  {journal}
		{\bibinfo  {journal} {Separation and Purification Technology}\ }\textbf
		{\bibinfo {volume} {251}},\ \bibinfo {pages} {117286} (\bibinfo {year}
		{2020})}\BibitemShut {NoStop}%
	\bibitem [{\citenamefont {Kim}\ \emph {et~al.}(2009)\citenamefont {Kim},
		\citenamefont {Ahn}, \citenamefont {Seok}, \citenamefont {Fleury},
		\citenamefont {Chang}, \citenamefont {Kim}, \citenamefont {Cha},\ and\
		\citenamefont {Kim}}]{MJH09}%
	\BibitemOpen
	\bibfield  {author} {\bibinfo {author} {\bibfnamefont {M.~U.}\ \bibnamefont
			{Kim}}, \bibinfo {author} {\bibfnamefont {J.~P.}\ \bibnamefont {Ahn}},
		\bibinfo {author} {\bibfnamefont {H.~K.}\ \bibnamefont {Seok}}, \bibinfo
		{author} {\bibfnamefont {E.}~\bibnamefont {Fleury}}, \bibinfo {author}
		{\bibfnamefont {H.~J.}\ \bibnamefont {Chang}}, \bibinfo {author}
		{\bibfnamefont {D.~H.}\ \bibnamefont {Kim}}, \bibinfo {author} {\bibfnamefont
			{P.~R.}\ \bibnamefont {Cha}},\ and\ \bibinfo {author} {\bibfnamefont {Y.~C.}\
			\bibnamefont {Kim}},\ }\href {https://doi.org/10.1007/s12540-009-0193-6}
	{\bibfield  {journal} {\bibinfo  {journal} {Metals and Materials
				International}\ }\textbf {\bibinfo {volume} {15}},\ \bibinfo {pages} {193}
		(\bibinfo {year} {2009})}\BibitemShut {NoStop}%
	\bibitem [{\citenamefont {Ingebrigtsen}\ and\ \citenamefont
		{Tanaka}(2016)}]{TH16}%
	\BibitemOpen
	\bibfield  {author} {\bibinfo {author} {\bibfnamefont {T.~S.}\ \bibnamefont
			{Ingebrigtsen}}\ and\ \bibinfo {author} {\bibfnamefont {H.}~\bibnamefont
			{Tanaka}},\ }\href {https://doi.org/10.1021/acs.jpcb.6b05486} {\bibfield
		{journal} {\bibinfo  {journal} {The Journal of Physical Chemistry B}\
		}\textbf {\bibinfo {volume} {120}},\ \bibinfo {pages} {7704} (\bibinfo {year}
		{2016})}\BibitemShut {NoStop}%
	\bibitem [{\citenamefont {Singh}\ and\ \citenamefont {Rabin}(2019)}]{KY19}%
	\BibitemOpen
	\bibfield  {author} {\bibinfo {author} {\bibfnamefont {K.}~\bibnamefont
			{Singh}}\ and\ \bibinfo {author} {\bibfnamefont {Y.}~\bibnamefont {Rabin}},\
	}\href {https://doi.org/10.1103/PhysRevLett.123.035502} {\bibfield  {journal}
		{\bibinfo  {journal} {Phys. Rev. Lett.}\ }\textbf {\bibinfo {volume} {123}},\
		\bibinfo {pages} {035502} (\bibinfo {year} {2019})}\BibitemShut {NoStop}%
	\bibitem [{\citenamefont {Priya}\ and\ \citenamefont {Jaiswal}(2020)}]{MP20}%
	\BibitemOpen
	\bibfield  {author} {\bibinfo {author} {\bibfnamefont {M.}~\bibnamefont
			{Priya}}\ and\ \bibinfo {author} {\bibfnamefont {P.~K.}\ \bibnamefont
			{Jaiswal}},\ }\href {https://doi.org/10.1080/01411594.2020.1813287}
	{\bibfield  {journal} {\bibinfo  {journal} {Phase Transitions}\ }\textbf
		{\bibinfo {volume} {93}},\ \bibinfo {pages} {895} (\bibinfo {year}
		{2020})}\BibitemShut {NoStop}%
	\bibitem [{\citenamefont {Lifshitz}\ and\ \citenamefont
		{Slyozov}(1961)}]{IV61}%
	\BibitemOpen
	\bibfield  {author} {\bibinfo {author} {\bibfnamefont {I.}~\bibnamefont
			{Lifshitz}}\ and\ \bibinfo {author} {\bibfnamefont {V.}~\bibnamefont
			{Slyozov}},\ }\href {https://doi.org/10.1016/0022-3697(61)90054-3} {\bibfield
		{journal} {\bibinfo  {journal} {Journal of Physics and Chemistry of Solids}\
		}\textbf {\bibinfo {volume} {19}},\ \bibinfo {pages} {35} (\bibinfo {year}
		{1961})}\BibitemShut {NoStop}%
	\bibitem [{\citenamefont {Kendon}\ \emph {et~al.}(2001)\citenamefont {Kendon},
		\citenamefont {Cates}, \citenamefont {Pagonabarraga}, \citenamefont
		{Desplat},\ and\ \citenamefont {Bladon}}]{VMI01}%
	\BibitemOpen
	\bibfield  {author} {\bibinfo {author} {\bibfnamefont {V.~M.}\ \bibnamefont
			{Kendon}}, \bibinfo {author} {\bibfnamefont {M.~E.}\ \bibnamefont {Cates}},
		\bibinfo {author} {\bibfnamefont {I.}~\bibnamefont {Pagonabarraga}}, \bibinfo
		{author} {\bibfnamefont {J.-C.}\ \bibnamefont {Desplat}},\ and\ \bibinfo
		{author} {\bibfnamefont {P.}~\bibnamefont {Bladon}},\ }\href
	{https://doi.org/10.1017/S0022112001004682} {\bibfield  {journal} {\bibinfo
			{journal} {Journal of Fluid Mechanics}\ }\textbf {\bibinfo {volume} {440}},\
		\bibinfo {pages} {147–203} (\bibinfo {year} {2001})}\BibitemShut {NoStop}%
	\bibitem [{\citenamefont {Wagner}\ and\ \citenamefont {Cates}(2001)}]{AM01}%
	\BibitemOpen
	\bibfield  {author} {\bibinfo {author} {\bibfnamefont {A.~J.}\ \bibnamefont
			{Wagner}}\ and\ \bibinfo {author} {\bibfnamefont {M.~E.}\ \bibnamefont
			{Cates}},\ }\href {https://doi.org/10.1209/epl/i2001-00551-4} {\bibfield
		{journal} {\bibinfo  {journal} {Europhysics Letters ({EPL})}\ }\textbf
		{\bibinfo {volume} {56}},\ \bibinfo {pages} {556} (\bibinfo {year}
		{2001})}\BibitemShut {NoStop}%
	\bibitem [{\citenamefont {Ahmad}\ \emph {et~al.}(2010)\citenamefont {Ahmad},
		\citenamefont {Das},\ and\ \citenamefont {Puri}}]{SSS10}%
	\BibitemOpen
	\bibfield  {author} {\bibinfo {author} {\bibfnamefont {S.}~\bibnamefont
			{Ahmad}}, \bibinfo {author} {\bibfnamefont {S.~K.}\ \bibnamefont {Das}},\
		and\ \bibinfo {author} {\bibfnamefont {S.}~\bibnamefont {Puri}},\ }\href
	{https://doi.org/10.1103/PhysRevE.82.040107} {\bibfield  {journal} {\bibinfo
			{journal} {Phys. Rev. E}\ }\textbf {\bibinfo {volume} {82}},\ \bibinfo
		{pages} {040107} (\bibinfo {year} {2010})}\BibitemShut {NoStop}%
	\bibitem [{\citenamefont {Ahmad}\ \emph {et~al.}(2012)\citenamefont {Ahmad},
		\citenamefont {Das},\ and\ \citenamefont {Puri}}]{SSS12}%
	\BibitemOpen
	\bibfield  {author} {\bibinfo {author} {\bibfnamefont {S.}~\bibnamefont
			{Ahmad}}, \bibinfo {author} {\bibfnamefont {S.~K.}\ \bibnamefont {Das}},\
		and\ \bibinfo {author} {\bibfnamefont {S.}~\bibnamefont {Puri}},\ }\href
	{https://doi.org/10.1103/PhysRevE.85.031140} {\bibfield  {journal} {\bibinfo
			{journal} {Phys. Rev. E}\ }\textbf {\bibinfo {volume} {85}},\ \bibinfo
		{pages} {031140} (\bibinfo {year} {2012})}\BibitemShut {NoStop}%
	\bibitem [{\citenamefont {Puri}(2005)}]{puri05}%
	\BibitemOpen
	\bibfield  {author} {\bibinfo {author} {\bibfnamefont {S.}~\bibnamefont
			{Puri}},\ }\href {https://doi.org/10.1088/0953-8984/17/3/r01} {\bibfield
		{journal} {\bibinfo  {journal} {Journal of Physics: Condensed Matter}\
		}\textbf {\bibinfo {volume} {17}},\ \bibinfo {pages} {R101} (\bibinfo {year}
		{2005})}\BibitemShut {NoStop}%
	\bibitem [{\citenamefont {Binder}\ \emph {et~al.}(2010)\citenamefont {Binder},
		\citenamefont {Puri}, \citenamefont {Das},\ and\ \citenamefont
		{Horbach}}]{KSS10}%
	\BibitemOpen
	\bibfield  {author} {\bibinfo {author} {\bibfnamefont {K.}~\bibnamefont
			{Binder}}, \bibinfo {author} {\bibfnamefont {S.}~\bibnamefont {Puri}},
		\bibinfo {author} {\bibfnamefont {S.~K.}\ \bibnamefont {Das}},\ and\ \bibinfo
		{author} {\bibfnamefont {J.}~\bibnamefont {Horbach}},\ }\href
	{https://doi.org/10.1007/s10955-010-9924-9} {\bibfield  {journal} {\bibinfo
			{journal} {Journal of Statistical Physics}\ }\textbf {\bibinfo {volume}
			{138}},\ \bibinfo {pages} {51} (\bibinfo {year} {2010})}\BibitemShut
	{NoStop}%
	\bibitem [{\citenamefont {van Franeker}\ \emph {et~al.}(2015)\citenamefont {van
			Franeker}, \citenamefont {Westhoff}, \citenamefont {Turbiez}, \citenamefont
		{Wienk}, \citenamefont {Schmidt},\ and\ \citenamefont {Janssen}}]{JDM15}%
	\BibitemOpen
	\bibfield  {author} {\bibinfo {author} {\bibfnamefont {J.~J.}\ \bibnamefont
			{van Franeker}}, \bibinfo {author} {\bibfnamefont {D.}~\bibnamefont
			{Westhoff}}, \bibinfo {author} {\bibfnamefont {M.}~\bibnamefont {Turbiez}},
		\bibinfo {author} {\bibfnamefont {M.~M.}\ \bibnamefont {Wienk}}, \bibinfo
		{author} {\bibfnamefont {V.}~\bibnamefont {Schmidt}},\ and\ \bibinfo {author}
		{\bibfnamefont {R.~A.~J.}\ \bibnamefont {Janssen}},\ }\href
	{https://doi.org/10.1002/adfm.201403392} {\bibfield  {journal} {\bibinfo
			{journal} {Advanced Functional Materials}\ }\textbf {\bibinfo {volume}
			{25}},\ \bibinfo {pages} {855} (\bibinfo {year} {2015})}\BibitemShut
	{NoStop}%
	\bibitem [{\citenamefont {Walheim}\ \emph {et~al.}(1997)\citenamefont
		{Walheim}, \citenamefont {Böltau}, \citenamefont {Mlynek}, \citenamefont
		{Krausch},\ and\ \citenamefont {Steiner}}]{SMJ97}%
	\BibitemOpen
	\bibfield  {author} {\bibinfo {author} {\bibfnamefont {S.}~\bibnamefont
			{Walheim}}, \bibinfo {author} {\bibfnamefont {M.}~\bibnamefont {Böltau}},
		\bibinfo {author} {\bibfnamefont {J.}~\bibnamefont {Mlynek}}, \bibinfo
		{author} {\bibfnamefont {G.}~\bibnamefont {Krausch}},\ and\ \bibinfo {author}
		{\bibfnamefont {U.}~\bibnamefont {Steiner}},\ }\href
	{https://doi.org/10.1021/ma9619288} {\bibfield  {journal} {\bibinfo
			{journal} {Macromolecules}\ }\textbf {\bibinfo {volume} {30}},\ \bibinfo
		{pages} {4995} (\bibinfo {year} {1997})}\BibitemShut {NoStop}%
	\bibitem [{\citenamefont {Michels}(2011)}]{Michels11}%
	\BibitemOpen
	\bibfield  {author} {\bibinfo {author} {\bibfnamefont {J.~J.}\ \bibnamefont
			{Michels}},\ }\href {https://doi.org/10.1002/cphc.201000873} {\bibfield
		{journal} {\bibinfo  {journal} {ChemPhysChem}\ }\textbf {\bibinfo {volume}
			{12}},\ \bibinfo {pages} {342} (\bibinfo {year} {2011})}\BibitemShut
	{NoStop}%
	\bibitem [{\citenamefont {Lambert}\ \emph {et~al.}(2010)\citenamefont
		{Lambert}, \citenamefont {Mokso}, \citenamefont {Cantat}, \citenamefont
		{Cloetens}, \citenamefont {Glazier}, \citenamefont {Graner},\ and\
		\citenamefont {Delannay}}]{JRI10}%
	\BibitemOpen
	\bibfield  {author} {\bibinfo {author} {\bibfnamefont {J.}~\bibnamefont
			{Lambert}}, \bibinfo {author} {\bibfnamefont {R.}~\bibnamefont {Mokso}},
		\bibinfo {author} {\bibfnamefont {I.}~\bibnamefont {Cantat}}, \bibinfo
		{author} {\bibfnamefont {P.}~\bibnamefont {Cloetens}}, \bibinfo {author}
		{\bibfnamefont {J.~A.}\ \bibnamefont {Glazier}}, \bibinfo {author}
		{\bibfnamefont {F.}~\bibnamefont {Graner}},\ and\ \bibinfo {author}
		{\bibfnamefont {R.}~\bibnamefont {Delannay}},\ }\href
	{https://doi.org/10.1103/PhysRevLett.104.248304} {\bibfield  {journal}
		{\bibinfo  {journal} {Phys. Rev. Lett.}\ }\textbf {\bibinfo {volume} {104}},\
		\bibinfo {pages} {248304} (\bibinfo {year} {2010})}\BibitemShut {NoStop}%
	\bibitem [{\citenamefont {Asadi}\ \emph {et~al.}(2008)\citenamefont {Asadi},
		\citenamefont {de~Leeuw}, \citenamefont {de~Boer},\ and\ \citenamefont
		{Blom}}]{KDB08}%
	\BibitemOpen
	\bibfield  {author} {\bibinfo {author} {\bibfnamefont {K.}~\bibnamefont
			{Asadi}}, \bibinfo {author} {\bibfnamefont {D.~M.}\ \bibnamefont {de~Leeuw}},
		\bibinfo {author} {\bibfnamefont {B.}~\bibnamefont {de~Boer}},\ and\ \bibinfo
		{author} {\bibfnamefont {P.~W.~M.}\ \bibnamefont {Blom}},\ }\href
	{https://doi.org/10.1038/nmat2207} {\bibfield  {journal} {\bibinfo  {journal}
			{Nature Materials}\ }\textbf {\bibinfo {volume} {7}},\ \bibinfo {pages} {547}
		(\bibinfo {year} {2008})}\BibitemShut {NoStop}%
	\bibitem [{\citenamefont {Chen}(2019)}]{Chen19}%
	\BibitemOpen
	\bibfield  {author} {\bibinfo {author} {\bibfnamefont {L.~X.}\ \bibnamefont
			{Chen}},\ }\href {https://doi.org/10.1021/acsenergylett.9b02071} {\bibfield
		{journal} {\bibinfo  {journal} {ACS Energy Letters}\ }\textbf {\bibinfo
			{volume} {4}},\ \bibinfo {pages} {2537} (\bibinfo {year} {2019})}\BibitemShut
	{NoStop}%
	\bibitem [{\citenamefont {Vaynzof}\ \emph {et~al.}(2011)\citenamefont
		{Vaynzof}, \citenamefont {Kabra}, \citenamefont {Zhao}, \citenamefont {Chua},
		\citenamefont {Steiner},\ and\ \citenamefont {Friend}}]{YDL11}%
	\BibitemOpen
	\bibfield  {author} {\bibinfo {author} {\bibfnamefont {Y.}~\bibnamefont
			{Vaynzof}}, \bibinfo {author} {\bibfnamefont {D.}~\bibnamefont {Kabra}},
		\bibinfo {author} {\bibfnamefont {L.}~\bibnamefont {Zhao}}, \bibinfo {author}
		{\bibfnamefont {L.~L.}\ \bibnamefont {Chua}}, \bibinfo {author}
		{\bibfnamefont {U.}~\bibnamefont {Steiner}},\ and\ \bibinfo {author}
		{\bibfnamefont {R.~H.}\ \bibnamefont {Friend}},\ }\href
	{https://doi.org/10.1021/nn102899g} {\bibfield  {journal} {\bibinfo
			{journal} {ACS Nano}\ }\textbf {\bibinfo {volume} {5}},\ \bibinfo {pages}
		{329} (\bibinfo {year} {2011})}\BibitemShut {NoStop}%
	\bibitem [{\citenamefont {Zhang}\ \emph {et~al.}(2020)\citenamefont {Zhang},
		\citenamefont {Li}, \citenamefont {Zhang}, \citenamefont {Zhang},\ and\
		\citenamefont {Zhou}}]{HYX20}%
	\BibitemOpen
	\bibfield  {author} {\bibinfo {author} {\bibfnamefont {H.}~\bibnamefont
			{Zhang}}, \bibinfo {author} {\bibfnamefont {Y.}~\bibnamefont {Li}}, \bibinfo
		{author} {\bibfnamefont {X.}~\bibnamefont {Zhang}}, \bibinfo {author}
		{\bibfnamefont {Y.}~\bibnamefont {Zhang}},\ and\ \bibinfo {author}
		{\bibfnamefont {H.}~\bibnamefont {Zhou}},\ }\href
	{https://doi.org/10.1039/D0QM00398K} {\bibfield  {journal} {\bibinfo
			{journal} {Mater. Chem. Front.}\ }\textbf {\bibinfo {volume} {4}},\ \bibinfo
		{pages} {2863} (\bibinfo {year} {2020})}\BibitemShut {NoStop}%
	\bibitem [{\citenamefont {Schaefer}\ \emph {et~al.}(2017)\citenamefont
		{Schaefer}, \citenamefont {Michels},\ and\ \citenamefont {van~der
			Schoot}}]{CJP17}%
	\BibitemOpen
	\bibfield  {author} {\bibinfo {author} {\bibfnamefont {C.}~\bibnamefont
			{Schaefer}}, \bibinfo {author} {\bibfnamefont {J.~J.}\ \bibnamefont
			{Michels}},\ and\ \bibinfo {author} {\bibfnamefont {P.}~\bibnamefont {van~der
				Schoot}},\ }\href {https://doi.org/10.1021/acs.macromol.7b01224} {\bibfield
		{journal} {\bibinfo  {journal} {Macromolecules}\ }\textbf {\bibinfo {volume}
			{50}},\ \bibinfo {pages} {5914} (\bibinfo {year} {2017})}\BibitemShut
	{NoStop}%
	\bibitem [{\citenamefont {Tabatabaieyazdi}\ \emph {et~al.}(2016)\citenamefont
		{Tabatabaieyazdi}, \citenamefont {Chan},\ and\ \citenamefont {Wu}}]{MPJ16}%
	\BibitemOpen
	\bibfield  {author} {\bibinfo {author} {\bibfnamefont {M.}~\bibnamefont
			{Tabatabaieyazdi}}, \bibinfo {author} {\bibfnamefont {P.~K.}\ \bibnamefont
			{Chan}},\ and\ \bibinfo {author} {\bibfnamefont {J.}~\bibnamefont {Wu}},\
	}\href {https://doi.org/10.1016/j.commatsci.2015.09.059} {\bibfield
		{journal} {\bibinfo  {journal} {Computational Materials Science}\ }\textbf
		{\bibinfo {volume} {111}},\ \bibinfo {pages} {387} (\bibinfo {year}
		{2016})}\BibitemShut {NoStop}%
	\bibitem [{\citenamefont {Epps{,}~III}\ and\ \citenamefont
		{O{'}Reilly}(2016)}]{TR16}%
	\BibitemOpen
	\bibfield  {author} {\bibinfo {author} {\bibfnamefont {T.~H.}\ \bibnamefont
			{Epps{,}~III}}\ and\ \bibinfo {author} {\bibfnamefont {R.~K.}\ \bibnamefont
			{O{'}Reilly}},\ }\href {https://doi.org/10.1039/C5SC03505H} {\bibfield
		{journal} {\bibinfo  {journal} {Chem. Sci.}\ }\textbf {\bibinfo {volume}
			{7}},\ \bibinfo {pages} {1674} (\bibinfo {year} {2016})}\BibitemShut
	{NoStop}%
	\bibitem [{\citenamefont {Das}\ \emph {et~al.}(2020{\natexlab{a}})\citenamefont
		{Das}, \citenamefont {Jaiswal},\ and\ \citenamefont {Puri}}]{PPS20}%
	\BibitemOpen
	\bibfield  {author} {\bibinfo {author} {\bibfnamefont {P.}~\bibnamefont
			{Das}}, \bibinfo {author} {\bibfnamefont {P.~K.}\ \bibnamefont {Jaiswal}},\
		and\ \bibinfo {author} {\bibfnamefont {S.}~\bibnamefont {Puri}},\ }\href
	{https://doi.org/10.1103/PhysRevE.102.012803} {\bibfield  {journal} {\bibinfo
			{journal} {Phys. Rev. E}\ }\textbf {\bibinfo {volume} {102}},\ \bibinfo
		{pages} {012803} (\bibinfo {year} {2020}{\natexlab{a}})}\BibitemShut
	{NoStop}%
	\bibitem [{\citenamefont {Das}\ \emph {et~al.}(2020{\natexlab{b}})\citenamefont
		{Das}, \citenamefont {Jaiswal},\ and\ \citenamefont {Puri}}]{PPS20-2}%
	\BibitemOpen
	\bibfield  {author} {\bibinfo {author} {\bibfnamefont {P.}~\bibnamefont
			{Das}}, \bibinfo {author} {\bibfnamefont {P.~K.}\ \bibnamefont {Jaiswal}},\
		and\ \bibinfo {author} {\bibfnamefont {S.}~\bibnamefont {Puri}},\ }\href
	{https://doi.org/10.1103/PhysRevE.102.032801} {\bibfield  {journal} {\bibinfo
			{journal} {Phys. Rev. E}\ }\textbf {\bibinfo {volume} {102}},\ \bibinfo
		{pages} {032801} (\bibinfo {year} {2020}{\natexlab{b}})}\BibitemShut
	{NoStop}%
	\bibitem [{\citenamefont {Ghosh}\ \emph {et~al.}(2020)\citenamefont {Ghosh},
		\citenamefont {Mukherjee}, \citenamefont {Arroyave},\ and\ \citenamefont
		{Douglas}}]{SAR20}%
	\BibitemOpen
	\bibfield  {author} {\bibinfo {author} {\bibfnamefont {S.}~\bibnamefont
			{Ghosh}}, \bibinfo {author} {\bibfnamefont {A.}~\bibnamefont {Mukherjee}},
		\bibinfo {author} {\bibfnamefont {R.}~\bibnamefont {Arroyave}},\ and\
		\bibinfo {author} {\bibfnamefont {J.~F.}\ \bibnamefont {Douglas}},\ }\href
	{https://doi.org/10.1063/5.0007859} {\bibfield  {journal} {\bibinfo
			{journal} {The Journal of Chemical Physics}\ }\textbf {\bibinfo {volume}
			{152}},\ \bibinfo {pages} {224902} (\bibinfo {year} {2020})}\BibitemShut
	{NoStop}%
	\bibitem [{\citenamefont {Cummins}\ \emph {et~al.}(2016)\citenamefont
		{Cummins}, \citenamefont {Ghoshal}, \citenamefont {Holmes},\ and\
		\citenamefont {Morris}}]{CTJ16}%
	\BibitemOpen
	\bibfield  {author} {\bibinfo {author} {\bibfnamefont {C.}~\bibnamefont
			{Cummins}}, \bibinfo {author} {\bibfnamefont {T.}~\bibnamefont {Ghoshal}},
		\bibinfo {author} {\bibfnamefont {J.~D.}\ \bibnamefont {Holmes}},\ and\
		\bibinfo {author} {\bibfnamefont {M.~A.}\ \bibnamefont {Morris}},\ }\href
	{https://doi.org/10.1002/adma.201503432} {\bibfield  {journal} {\bibinfo
			{journal} {Advanced Materials}\ }\textbf {\bibinfo {volume} {28}},\ \bibinfo
		{pages} {5586} (\bibinfo {year} {2016})}\BibitemShut {NoStop}%
	\bibitem [{\citenamefont {Majewski}\ \emph {et~al.}(2015)\citenamefont
		{Majewski}, \citenamefont {Rahman}, \citenamefont {Black},\ and\
		\citenamefont {Yager}}]{PAC15}%
	\BibitemOpen
	\bibfield  {author} {\bibinfo {author} {\bibfnamefont {P.~W.}\ \bibnamefont
			{Majewski}}, \bibinfo {author} {\bibfnamefont {A.}~\bibnamefont {Rahman}},
		\bibinfo {author} {\bibfnamefont {C.~T.}\ \bibnamefont {Black}},\ and\
		\bibinfo {author} {\bibfnamefont {K.~G.}\ \bibnamefont {Yager}},\ }\href
	{https://doi.org/10.1038/ncomms8448} {\bibfield  {journal} {\bibinfo
			{journal} {Nature Communications}\ }\textbf {\bibinfo {volume} {6}},\
		\bibinfo {pages} {7448} (\bibinfo {year} {2015})}\BibitemShut {NoStop}%
	\bibitem [{\citenamefont {Kim}\ \emph {et~al.}(2014)\citenamefont {Kim},
		\citenamefont {Gwyther}, \citenamefont {Manners}, \citenamefont {Chaikin},\
		and\ \citenamefont {Register}}]{SJI14}%
	\BibitemOpen
	\bibfield  {author} {\bibinfo {author} {\bibfnamefont {S.~Y.}\ \bibnamefont
			{Kim}}, \bibinfo {author} {\bibfnamefont {J.}~\bibnamefont {Gwyther}},
		\bibinfo {author} {\bibfnamefont {I.}~\bibnamefont {Manners}}, \bibinfo
		{author} {\bibfnamefont {P.~M.}\ \bibnamefont {Chaikin}},\ and\ \bibinfo
		{author} {\bibfnamefont {R.~A.}\ \bibnamefont {Register}},\ }\href
	{https://doi.org/10.1002/adma.201303452} {\bibfield  {journal} {\bibinfo
			{journal} {Advanced Materials}\ }\textbf {\bibinfo {volume} {26}},\ \bibinfo
		{pages} {791} (\bibinfo {year} {2014})}\BibitemShut {NoStop}%
	\bibitem [{\citenamefont {Werner}\ \emph {et~al.}(2018)\citenamefont {Werner},
		\citenamefont {Rodríguez-Calero}, \citenamefont {Abruña},\ and\
		\citenamefont {Wiesner}}]{JGH18}%
	\BibitemOpen
	\bibfield  {author} {\bibinfo {author} {\bibfnamefont {J.~G.}\ \bibnamefont
			{Werner}}, \bibinfo {author} {\bibfnamefont {G.~G.}\ \bibnamefont
			{Rodríguez-Calero}}, \bibinfo {author} {\bibfnamefont {H.~D.}\ \bibnamefont
			{Abruña}},\ and\ \bibinfo {author} {\bibfnamefont {U.}~\bibnamefont
			{Wiesner}},\ }\href {https://doi.org/10.1039/C7EE03571C} {\bibfield
		{journal} {\bibinfo  {journal} {Energy Environ. Sci.}\ }\textbf {\bibinfo
			{volume} {11}},\ \bibinfo {pages} {1261} (\bibinfo {year}
		{2018})}\BibitemShut {NoStop}%
	\bibitem [{\citenamefont {Bates}\ \emph {et~al.}(2014)\citenamefont {Bates},
		\citenamefont {Maher}, \citenamefont {Janes}, \citenamefont {Ellison},\ and\
		\citenamefont {Willson}}]{CMD14}%
	\BibitemOpen
	\bibfield  {author} {\bibinfo {author} {\bibfnamefont {C.~M.}\ \bibnamefont
			{Bates}}, \bibinfo {author} {\bibfnamefont {M.~J.}\ \bibnamefont {Maher}},
		\bibinfo {author} {\bibfnamefont {D.~W.}\ \bibnamefont {Janes}}, \bibinfo
		{author} {\bibfnamefont {C.~J.}\ \bibnamefont {Ellison}},\ and\ \bibinfo
		{author} {\bibfnamefont {C.~G.}\ \bibnamefont {Willson}},\ }\href
	{https://doi.org/10.1021/ma401762n} {\bibfield  {journal} {\bibinfo
			{journal} {Macromolecules}\ }\textbf {\bibinfo {volume} {47}},\ \bibinfo
		{pages} {2} (\bibinfo {year} {2014})}\BibitemShut {NoStop}%
	\bibitem [{\citenamefont {Liu}\ and\ \citenamefont {Liu}(2019)}]{TG19}%
	\BibitemOpen
	\bibfield  {author} {\bibinfo {author} {\bibfnamefont {T.}~\bibnamefont
			{Liu}}\ and\ \bibinfo {author} {\bibfnamefont {G.}~\bibnamefont {Liu}},\
	}\href {https://doi.org/10.1088/1361-648x/ab0d77} {\bibfield  {journal}
		{\bibinfo  {journal} {Journal of Physics: Condensed Matter}\ }\textbf
		{\bibinfo {volume} {31}},\ \bibinfo {pages} {233001} (\bibinfo {year}
		{2019})}\BibitemShut {NoStop}%
	\bibitem [{\citenamefont {Carmona}\ \emph {et~al.}(2021)\citenamefont
		{Carmona}, \citenamefont {R{\"o}ding}, \citenamefont {S{\"a}rkk{\"a}},
		\citenamefont {von Corswant}, \citenamefont {Olsson},\ and\ \citenamefont
		{Lor{\'e}n}}]{PMA21}%
	\BibitemOpen
	\bibfield  {author} {\bibinfo {author} {\bibfnamefont {P.}~\bibnamefont
			{Carmona}}, \bibinfo {author} {\bibfnamefont {M.}~\bibnamefont {R{\"o}ding}},
		\bibinfo {author} {\bibfnamefont {A.}~\bibnamefont {S{\"a}rkk{\"a}}},
		\bibinfo {author} {\bibfnamefont {C.}~\bibnamefont {von Corswant}}, \bibinfo
		{author} {\bibfnamefont {E.}~\bibnamefont {Olsson}},\ and\ \bibinfo {author}
		{\bibfnamefont {N.}~\bibnamefont {Lor{\'e}n}},\ }\href
	{https://doi.org/10.1039/D1SM00044F} {\bibfield  {journal} {\bibinfo
			{journal} {Soft Matter}\ }\textbf {\bibinfo {volume} {17}},\ \bibinfo {pages}
		{3913} (\bibinfo {year} {2021})}\BibitemShut {NoStop}%
	\bibitem [{\citenamefont {Pascual}\ \emph {et~al.}(2021)\citenamefont
		{Pascual}, \citenamefont {Amon},\ and\ \citenamefont {Jullien}}]{MAM21}%
	\BibitemOpen
	\bibfield  {author} {\bibinfo {author} {\bibfnamefont {M.}~\bibnamefont
			{Pascual}}, \bibinfo {author} {\bibfnamefont {A.}~\bibnamefont {Amon}},\ and\
		\bibinfo {author} {\bibfnamefont {M.-C.}\ \bibnamefont {Jullien}},\ }\href
	{https://doi.org/10.1103/PhysRevFluids.6.114203} {\bibfield  {journal}
		{\bibinfo  {journal} {Phys. Rev. Fluids}\ }\textbf {\bibinfo {volume} {6}},\
		\bibinfo {pages} {114203} (\bibinfo {year} {2021})}\BibitemShut {NoStop}%
	\bibitem [{\citenamefont {Jamie}\ \emph {et~al.}(2012)\citenamefont {Jamie},
		\citenamefont {Dullens},\ and\ \citenamefont {Aarts}}]{ERD12}%
	\BibitemOpen
	\bibfield  {author} {\bibinfo {author} {\bibfnamefont {E.~A.~G.}\
			\bibnamefont {Jamie}}, \bibinfo {author} {\bibfnamefont {R.~P.~A.}\
			\bibnamefont {Dullens}},\ and\ \bibinfo {author} {\bibfnamefont {D.~G.
				A.~L.}\ \bibnamefont {Aarts}},\ }\href {https://doi.org/10.1063/1.4767399}
	{\bibfield  {journal} {\bibinfo  {journal} {The Journal of Chemical Physics}\
		}\textbf {\bibinfo {volume} {137}},\ \bibinfo {pages} {204902} (\bibinfo
		{year} {2012})}\BibitemShut {NoStop}%
	\bibitem [{\citenamefont {Puri}\ and\ \citenamefont
		{Binder}(1992{\natexlab{a}})}]{SK92}%
	\BibitemOpen
	\bibfield  {author} {\bibinfo {author} {\bibfnamefont {S.}~\bibnamefont
			{Puri}}\ and\ \bibinfo {author} {\bibfnamefont {K.}~\bibnamefont {Binder}},\
	}\href {https://doi.org/10.1103/PhysRevA.46.R4487} {\bibfield  {journal}
		{\bibinfo  {journal} {Phys. Rev. A}\ }\textbf {\bibinfo {volume} {46}},\
		\bibinfo {pages} {R4487} (\bibinfo {year} {1992}{\natexlab{a}})}\BibitemShut
	{NoStop}%
	\bibitem [{\citenamefont {Troian}(1993)}]{troian93prl}%
	\BibitemOpen
	\bibfield  {author} {\bibinfo {author} {\bibfnamefont {S.~M.}\ \bibnamefont
			{Troian}},\ }\href {https://doi.org/10.1103/PhysRevLett.71.1399} {\bibfield
		{journal} {\bibinfo  {journal} {Phys. Rev. Lett.}\ }\textbf {\bibinfo
			{volume} {71}},\ \bibinfo {pages} {1399} (\bibinfo {year}
		{1993})}\BibitemShut {NoStop}%
	\bibitem [{\citenamefont {Troian}(1992)}]{troian92}%
	\BibitemOpen
	\bibfield  {author} {\bibinfo {author} {\bibfnamefont {S.~M.}\ \bibnamefont
			{Troian}},\ }\href {https://doi.org/10.1557/PROC-290-29} {\bibfield
		{journal} {\bibinfo  {journal} {MRS Online Proceedings Library}\ }\textbf
		{\bibinfo {volume} {290}},\ \bibinfo {pages} {29} (\bibinfo {year}
		{1992})}\BibitemShut {NoStop}%
	\bibitem [{\citenamefont {Fischer}\ \emph {et~al.}(1997)\citenamefont
		{Fischer}, \citenamefont {Maass},\ and\ \citenamefont {Dieterich}}]{HPW97}%
	\BibitemOpen
	\bibfield  {author} {\bibinfo {author} {\bibfnamefont {H.~P.}\ \bibnamefont
			{Fischer}}, \bibinfo {author} {\bibfnamefont {P.}~\bibnamefont {Maass}},\
		and\ \bibinfo {author} {\bibfnamefont {W.}~\bibnamefont {Dieterich}},\ }\href
	{https://doi.org/10.1103/PhysRevLett.79.893} {\bibfield  {journal} {\bibinfo
			{journal} {Phys. Rev. Lett.}\ }\textbf {\bibinfo {volume} {79}},\ \bibinfo
		{pages} {893} (\bibinfo {year} {1997})}\BibitemShut {NoStop}%
	\bibitem [{\citenamefont {Ball}\ and\ \citenamefont {Essery}(1990)}]{RR90}%
	\BibitemOpen
	\bibfield  {author} {\bibinfo {author} {\bibfnamefont {R.~C.}\ \bibnamefont
			{Ball}}\ and\ \bibinfo {author} {\bibfnamefont {R.~L.~H.}\ \bibnamefont
			{Essery}},\ }\href {https://doi.org/10.1088/0953-8984/2/51/006} {\bibfield
		{journal} {\bibinfo  {journal} {Journal of Physics: Condensed Matter}\
		}\textbf {\bibinfo {volume} {2}},\ \bibinfo {pages} {10303} (\bibinfo {year}
		{1990})}\BibitemShut {NoStop}%
	\bibitem [{\citenamefont {Binder}\ and\ \citenamefont {Frisch}(1991)}]{KH91}%
	\BibitemOpen
	\bibfield  {author} {\bibinfo {author} {\bibfnamefont {K.}~\bibnamefont
			{Binder}}\ and\ \bibinfo {author} {\bibfnamefont {H.~L.}\ \bibnamefont
			{Frisch}},\ }\href {https://doi.org/10.1007/BF01314015} {\bibfield  {journal}
		{\bibinfo  {journal} {Zeitschrift f{\"u}r Physik B Condensed Matter}\
		}\textbf {\bibinfo {volume} {84}},\ \bibinfo {pages} {403} (\bibinfo {year}
		{1991})}\BibitemShut {NoStop}%
	\bibitem [{\citenamefont {Marko}(1993)}]{marko93}%
	\BibitemOpen
	\bibfield  {author} {\bibinfo {author} {\bibfnamefont {J.~F.}\ \bibnamefont
			{Marko}},\ }\href {https://doi.org/10.1103/PhysRevE.48.2861} {\bibfield
		{journal} {\bibinfo  {journal} {Phys. Rev. E}\ }\textbf {\bibinfo {volume}
			{48}},\ \bibinfo {pages} {2861} (\bibinfo {year} {1993})}\BibitemShut
	{NoStop}%
	\bibitem [{\citenamefont {Geoghegan}\ and\ \citenamefont
		{Krausch}(2003)}]{MG03}%
	\BibitemOpen
	\bibfield  {author} {\bibinfo {author} {\bibfnamefont {M.}~\bibnamefont
			{Geoghegan}}\ and\ \bibinfo {author} {\bibfnamefont {G.}~\bibnamefont
			{Krausch}},\ }\href {https://doi.org/10.1016/S0079-6700(02)00080-1}
	{\bibfield  {journal} {\bibinfo  {journal} {Progress in Polymer Science}\
		}\textbf {\bibinfo {volume} {28}},\ \bibinfo {pages} {261} (\bibinfo {year}
		{2003})}\BibitemShut {NoStop}%
	\bibitem [{\citenamefont {Krausch}(1995)}]{krausch95}%
	\BibitemOpen
	\bibfield  {author} {\bibinfo {author} {\bibfnamefont {G.}~\bibnamefont
			{Krausch}},\ }\href {https://doi.org/10.1016/0927-796X(94)00173-1} {\bibfield
		{journal} {\bibinfo  {journal} {Materials Science and Engineering: R:
				Reports}\ }\textbf {\bibinfo {volume} {14}},\ \bibinfo {pages} {v} (\bibinfo
		{year} {1995})}\BibitemShut {NoStop}%
	\bibitem [{\citenamefont {Wiltzius}\ and\ \citenamefont
		{Cumming}(1991)}]{PA91}%
	\BibitemOpen
	\bibfield  {author} {\bibinfo {author} {\bibfnamefont {P.}~\bibnamefont
			{Wiltzius}}\ and\ \bibinfo {author} {\bibfnamefont {A.}~\bibnamefont
			{Cumming}},\ }\href {https://doi.org/10.1103/PhysRevLett.66.3000} {\bibfield
		{journal} {\bibinfo  {journal} {Phys. Rev. Lett.}\ }\textbf {\bibinfo
			{volume} {66}},\ \bibinfo {pages} {3000} (\bibinfo {year}
		{1991})}\BibitemShut {NoStop}%
	\bibitem [{\citenamefont {Cumming}\ \emph {et~al.}(1992)\citenamefont
		{Cumming}, \citenamefont {Wiltzius}, \citenamefont {Bates},\ and\
		\citenamefont {Rosedale}}]{APF92}%
	\BibitemOpen
	\bibfield  {author} {\bibinfo {author} {\bibfnamefont {A.}~\bibnamefont
			{Cumming}}, \bibinfo {author} {\bibfnamefont {P.}~\bibnamefont {Wiltzius}},
		\bibinfo {author} {\bibfnamefont {F.~S.}\ \bibnamefont {Bates}},\ and\
		\bibinfo {author} {\bibfnamefont {J.~H.}\ \bibnamefont {Rosedale}},\ }\href
	{https://doi.org/10.1103/PhysRevA.45.885} {\bibfield  {journal} {\bibinfo
			{journal} {Phys. Rev. A}\ }\textbf {\bibinfo {volume} {45}},\ \bibinfo
		{pages} {885} (\bibinfo {year} {1992})}\BibitemShut {NoStop}%
	\bibitem [{\citenamefont {Shi}\ \emph {et~al.}(1993)\citenamefont {Shi},
		\citenamefont {Harrison},\ and\ \citenamefont {Cumming}}]{BCA93}%
	\BibitemOpen
	\bibfield  {author} {\bibinfo {author} {\bibfnamefont {B.~Q.}\ \bibnamefont
			{Shi}}, \bibinfo {author} {\bibfnamefont {C.}~\bibnamefont {Harrison}},\ and\
		\bibinfo {author} {\bibfnamefont {A.}~\bibnamefont {Cumming}},\ }\href
	{https://doi.org/10.1103/PhysRevLett.70.206} {\bibfield  {journal} {\bibinfo
			{journal} {Phys. Rev. Lett.}\ }\textbf {\bibinfo {volume} {70}},\ \bibinfo
		{pages} {206} (\bibinfo {year} {1993})}\BibitemShut {NoStop}%
	\bibitem [{\citenamefont {Puri}\ and\ \citenamefont {Oono}(1988)}]{SY88}%
	\BibitemOpen
	\bibfield  {author} {\bibinfo {author} {\bibfnamefont {S.}~\bibnamefont
			{Puri}}\ and\ \bibinfo {author} {\bibfnamefont {Y.}~\bibnamefont {Oono}},\
	}\href {https://doi.org/10.1088/0305-4470/21/15/003} {\bibfield  {journal}
		{\bibinfo  {journal} {Journal of Physics A: Mathematical and General}\
		}\textbf {\bibinfo {volume} {21}},\ \bibinfo {pages} {L755} (\bibinfo {year}
		{1988})}\BibitemShut {NoStop}%
	\bibitem [{\citenamefont {Puri}\ and\ \citenamefont {Frisch}(1997)}]{SH97}%
	\BibitemOpen
	\bibfield  {author} {\bibinfo {author} {\bibfnamefont {S.}~\bibnamefont
			{Puri}}\ and\ \bibinfo {author} {\bibfnamefont {H.~L.}\ \bibnamefont
			{Frisch}},\ }\href {https://doi.org/10.1088/0953-8984/9/10/003} {\bibfield
		{journal} {\bibinfo  {journal} {Journal of Physics: Condensed Matter}\
		}\textbf {\bibinfo {volume} {9}},\ \bibinfo {pages} {2109} (\bibinfo {year}
		{1997})}\BibitemShut {NoStop}%
	\bibitem [{\citenamefont {Binder}(1983)}]{binder83}%
	\BibitemOpen
	\bibfield  {author} {\bibinfo {author} {\bibfnamefont {K.}~\bibnamefont
			{Binder}},\ }\href {https://doi.org/10.1063/1.445747} {\bibfield  {journal}
		{\bibinfo  {journal} {The Journal of Chemical Physics}\ }\textbf {\bibinfo
			{volume} {79}},\ \bibinfo {pages} {6387} (\bibinfo {year}
		{1983})}\BibitemShut {NoStop}%
	\bibitem [{\citenamefont {Brown}\ and\ \citenamefont
		{Chakrabarti}(1992)}]{GA92}%
	\BibitemOpen
	\bibfield  {author} {\bibinfo {author} {\bibfnamefont {G.}~\bibnamefont
			{Brown}}\ and\ \bibinfo {author} {\bibfnamefont {A.}~\bibnamefont
			{Chakrabarti}},\ }\href {https://doi.org/10.1103/PhysRevA.46.4829} {\bibfield
		{journal} {\bibinfo  {journal} {Phys. Rev. A}\ }\textbf {\bibinfo {volume}
			{46}},\ \bibinfo {pages} {4829} (\bibinfo {year} {1992})}\BibitemShut
	{NoStop}%
	\bibitem [{\citenamefont {Puri}\ and\ \citenamefont {Binder}(2001)}]{SK01}%
	\BibitemOpen
	\bibfield  {author} {\bibinfo {author} {\bibfnamefont {S.}~\bibnamefont
			{Puri}}\ and\ \bibinfo {author} {\bibfnamefont {K.}~\bibnamefont {Binder}},\
	}\href {https://doi.org/10.1103/PhysRevLett.86.1797} {\bibfield  {journal}
		{\bibinfo  {journal} {Phys. Rev. Lett.}\ }\textbf {\bibinfo {volume} {86}},\
		\bibinfo {pages} {1797} (\bibinfo {year} {2001})}\BibitemShut {NoStop}%
	\bibitem [{\citenamefont {Puri}\ and\ \citenamefont {Binder}(1994)}]{SK94}%
	\BibitemOpen
	\bibfield  {author} {\bibinfo {author} {\bibfnamefont {S.}~\bibnamefont
			{Puri}}\ and\ \bibinfo {author} {\bibfnamefont {K.}~\bibnamefont {Binder}},\
	}\href {https://doi.org/10.1103/PhysRevE.49.5359} {\bibfield  {journal}
		{\bibinfo  {journal} {Phys. Rev. E}\ }\textbf {\bibinfo {volume} {49}},\
		\bibinfo {pages} {5359} (\bibinfo {year} {1994})}\BibitemShut {NoStop}%
	\bibitem [{\citenamefont {Jaiswal}\ \emph
		{et~al.}(2012{\natexlab{a}})\citenamefont {Jaiswal}, \citenamefont {Puri},\
		and\ \citenamefont {Das}}]{PSS12}%
	\BibitemOpen
	\bibfield  {author} {\bibinfo {author} {\bibfnamefont {P.~K.}\ \bibnamefont
			{Jaiswal}}, \bibinfo {author} {\bibfnamefont {S.}~\bibnamefont {Puri}},\ and\
		\bibinfo {author} {\bibfnamefont {S.~K.}\ \bibnamefont {Das}},\ }\href
	{https://doi.org/10.1103/PhysRevE.85.051137} {\bibfield  {journal} {\bibinfo
			{journal} {Phys. Rev. E}\ }\textbf {\bibinfo {volume} {85}},\ \bibinfo
		{pages} {051137} (\bibinfo {year} {2012}{\natexlab{a}})}\BibitemShut
	{NoStop}%
	\bibitem [{\citenamefont {Bastea}\ \emph {et~al.}(2001)\citenamefont {Bastea},
		\citenamefont {Puri},\ and\ \citenamefont {Lebowitz}}]{SSJ01}%
	\BibitemOpen
	\bibfield  {author} {\bibinfo {author} {\bibfnamefont {S.}~\bibnamefont
			{Bastea}}, \bibinfo {author} {\bibfnamefont {S.}~\bibnamefont {Puri}},\ and\
		\bibinfo {author} {\bibfnamefont {J.~L.}\ \bibnamefont {Lebowitz}},\ }\href
	{https://doi.org/10.1103/PhysRevE.63.041513} {\bibfield  {journal} {\bibinfo
			{journal} {Phys. Rev. E}\ }\textbf {\bibinfo {volume} {63}},\ \bibinfo
		{pages} {041513} (\bibinfo {year} {2001})}\BibitemShut {NoStop}%
	\bibitem [{\citenamefont {Das}\ \emph {et~al.}(2005)\citenamefont {Das},
		\citenamefont {Puri}, \citenamefont {Horbach},\ and\ \citenamefont
		{Binder}}]{SSJ05}%
	\BibitemOpen
	\bibfield  {author} {\bibinfo {author} {\bibfnamefont {S.~K.}\ \bibnamefont
			{Das}}, \bibinfo {author} {\bibfnamefont {S.}~\bibnamefont {Puri}}, \bibinfo
		{author} {\bibfnamefont {J.}~\bibnamefont {Horbach}},\ and\ \bibinfo {author}
		{\bibfnamefont {K.}~\bibnamefont {Binder}},\ }\href
	{https://doi.org/10.1103/PhysRevE.72.061603} {\bibfield  {journal} {\bibinfo
			{journal} {Phys. Rev. E}\ }\textbf {\bibinfo {volume} {72}},\ \bibinfo
		{pages} {061603} (\bibinfo {year} {2005})}\BibitemShut {NoStop}%
	\bibitem [{\citenamefont {Das}\ \emph {et~al.}(2006)\citenamefont {Das},
		\citenamefont {Puri}, \citenamefont {Horbach},\ and\ \citenamefont
		{Binder}}]{SSJ06-pre}%
	\BibitemOpen
	\bibfield  {author} {\bibinfo {author} {\bibfnamefont {S.~K.}\ \bibnamefont
			{Das}}, \bibinfo {author} {\bibfnamefont {S.}~\bibnamefont {Puri}}, \bibinfo
		{author} {\bibfnamefont {J.}~\bibnamefont {Horbach}},\ and\ \bibinfo {author}
		{\bibfnamefont {K.}~\bibnamefont {Binder}},\ }\href
	{https://doi.org/10.1103/PhysRevE.73.031604} {\bibfield  {journal} {\bibinfo
			{journal} {Phys. Rev. E}\ }\textbf {\bibinfo {volume} {73}},\ \bibinfo
		{pages} {031604} (\bibinfo {year} {2006})}\BibitemShut {NoStop}%
	\bibitem [{\citenamefont {Goyal}\ \emph {et~al.}(2021)\citenamefont {Goyal},
		\citenamefont {van~der Schoot},\ and\ \citenamefont {Toschi}}]{APF21}%
	\BibitemOpen
	\bibfield  {author} {\bibinfo {author} {\bibfnamefont {A.}~\bibnamefont
			{Goyal}}, \bibinfo {author} {\bibfnamefont {P.}~\bibnamefont {van~der
				Schoot}},\ and\ \bibinfo {author} {\bibfnamefont {F.}~\bibnamefont
			{Toschi}},\ }\href {https://doi.org/10.1103/PhysRevE.103.042801} {\bibfield
		{journal} {\bibinfo  {journal} {Phys. Rev. E}\ }\textbf {\bibinfo {volume}
			{103}},\ \bibinfo {pages} {042801} (\bibinfo {year} {2021})}\BibitemShut
	{NoStop}%
	\bibitem [{\citenamefont {Toxvaerd}(1999)}]{toxvaerd99prl}%
	\BibitemOpen
	\bibfield  {author} {\bibinfo {author} {\bibfnamefont {S.}~\bibnamefont
			{Toxvaerd}},\ }\href {https://doi.org/10.1103/PhysRevLett.83.5318} {\bibfield
		{journal} {\bibinfo  {journal} {Phys. Rev. Lett.}\ }\textbf {\bibinfo
			{volume} {83}},\ \bibinfo {pages} {5318} (\bibinfo {year}
		{1999})}\BibitemShut {NoStop}%
	\bibitem [{\citenamefont {Chen}\ and\ \citenamefont
		{Chakrabarti}(1997)}]{HA97pre}%
	\BibitemOpen
	\bibfield  {author} {\bibinfo {author} {\bibfnamefont {H.}~\bibnamefont
			{Chen}}\ and\ \bibinfo {author} {\bibfnamefont {A.}~\bibnamefont
			{Chakrabarti}},\ }\href {https://doi.org/10.1103/PhysRevE.55.5680} {\bibfield
		{journal} {\bibinfo  {journal} {Phys. Rev. E}\ }\textbf {\bibinfo {volume}
			{55}},\ \bibinfo {pages} {5680} (\bibinfo {year} {1997})}\BibitemShut
	{NoStop}%
	\bibitem [{\citenamefont {Keblinski}\ \emph {et~al.}(1994)\citenamefont
		{Keblinski}, \citenamefont {Ma}, \citenamefont {Maritan}, \citenamefont
		{Koplik},\ and\ \citenamefont {Banavar}}]{PWA94prl}%
	\BibitemOpen
	\bibfield  {author} {\bibinfo {author} {\bibfnamefont {P.}~\bibnamefont
			{Keblinski}}, \bibinfo {author} {\bibfnamefont {W.~J.}\ \bibnamefont {Ma}},
		\bibinfo {author} {\bibfnamefont {A.}~\bibnamefont {Maritan}}, \bibinfo
		{author} {\bibfnamefont {J.}~\bibnamefont {Koplik}},\ and\ \bibinfo {author}
		{\bibfnamefont {J.~R.}\ \bibnamefont {Banavar}},\ }\href
	{https://doi.org/10.1103/PhysRevLett.72.3738} {\bibfield  {journal} {\bibinfo
			{journal} {Phys. Rev. Lett.}\ }\textbf {\bibinfo {volume} {72}},\ \bibinfo
		{pages} {3738} (\bibinfo {year} {1994})}\BibitemShut {NoStop}%
	\bibitem [{\citenamefont {Tanaka}\ and\ \citenamefont {Araki}(2000)}]{HT00}%
	\BibitemOpen
	\bibfield  {author} {\bibinfo {author} {\bibfnamefont {H.}~\bibnamefont
			{Tanaka}}\ and\ \bibinfo {author} {\bibfnamefont {T.}~\bibnamefont {Araki}},\
	}\href {https://doi.org/10.1209/epl/i2000-00525-0} {\bibfield  {journal}
		{\bibinfo  {journal} {Europhysics Letters ({EPL})}\ }\textbf {\bibinfo
			{volume} {51}},\ \bibinfo {pages} {154} (\bibinfo {year} {2000})}\BibitemShut
	{NoStop}%
	\bibitem [{\citenamefont {Tanaka}(2001)}]{tanaka01}%
	\BibitemOpen
	\bibfield  {author} {\bibinfo {author} {\bibfnamefont {H.}~\bibnamefont
			{Tanaka}},\ }\href {https://doi.org/10.1088/0953-8984/13/21/303} {\bibfield
		{journal} {\bibinfo  {journal} {Journal of Physics: Condensed Matter}\
		}\textbf {\bibinfo {volume} {13}},\ \bibinfo {pages} {4637} (\bibinfo {year}
		{2001})}\BibitemShut {NoStop}%
	\bibitem [{\citenamefont {Puri}\ and\ \citenamefont
		{Binder}(1992{\natexlab{b}})}]{SK92-1}%
	\BibitemOpen
	\bibfield  {author} {\bibinfo {author} {\bibfnamefont {S.}~\bibnamefont
			{Puri}}\ and\ \bibinfo {author} {\bibfnamefont {K.}~\bibnamefont {Binder}},\
	}\href {https://doi.org/10.1007/BF01313835} {\bibfield  {journal} {\bibinfo
			{journal} {Zeitschrift f{\"u}r Physik B Condensed Matter}\ }\textbf {\bibinfo
			{volume} {86}},\ \bibinfo {pages} {263} (\bibinfo {year}
		{1992}{\natexlab{b}})}\BibitemShut {NoStop}%
	\bibitem [{\citenamefont {Plimpton}(1995)}]{plimpton95}%
	\BibitemOpen
	\bibfield  {author} {\bibinfo {author} {\bibfnamefont {S.}~\bibnamefont
			{Plimpton}},\ }\href {https://doi.org/10.1006/jcph.1995.1039} {\bibfield
		{journal} {\bibinfo  {journal} {Journal of Computational Physics}\ }\textbf
		{\bibinfo {volume} {117}},\ \bibinfo {pages} {1} (\bibinfo {year}
		{1995})}\BibitemShut {NoStop}%
	\bibitem [{sm()}]{sm}%
	\BibitemOpen
	\href@noop {} {\bibinfo {title} {See supplemental material for the other
			details, snapshots and corresponding movies}}\BibitemShut {NoStop}%
	\bibitem [{\citenamefont {Das}\ and\ \citenamefont {Binder}(2010)}]{SK10}%
	\BibitemOpen
	\bibfield  {author} {\bibinfo {author} {\bibfnamefont {S.~K.}\ \bibnamefont
			{Das}}\ and\ \bibinfo {author} {\bibfnamefont {K.}~\bibnamefont {Binder}},\
	}\href {https://doi.org/10.1209/0295-5075/92/26006} {\bibfield  {journal}
		{\bibinfo  {journal} {{Europhysics Letters} (EPL)}\ }\textbf {\bibinfo
			{volume} {92}},\ \bibinfo {pages} {26006} (\bibinfo {year}
		{2010})}\BibitemShut {NoStop}%
	\bibitem [{\citenamefont {Tanaka}(1993{\natexlab{a}})}]{tanaka93-2}%
	\BibitemOpen
	\bibfield  {author} {\bibinfo {author} {\bibfnamefont {H.}~\bibnamefont
			{Tanaka}},\ }\href {https://doi.org/10.1103/PhysRevLett.70.53} {\bibfield
		{journal} {\bibinfo  {journal} {Phys. Rev. Lett.}\ }\textbf {\bibinfo
			{volume} {70}},\ \bibinfo {pages} {53} (\bibinfo {year}
		{1993}{\natexlab{a}})}\BibitemShut {NoStop}%
	\bibitem [{\citenamefont {Tanaka}(1993{\natexlab{b}})}]{tanaka93}%
	\BibitemOpen
	\bibfield  {author} {\bibinfo {author} {\bibfnamefont {H.}~\bibnamefont
			{Tanaka}},\ }\href {https://doi.org/10.1103/PhysRevLett.70.2770} {\bibfield
		{journal} {\bibinfo  {journal} {Phys. Rev. Lett.}\ }\textbf {\bibinfo
			{volume} {70}},\ \bibinfo {pages} {2770} (\bibinfo {year}
		{1993}{\natexlab{b}})}\BibitemShut {NoStop}%
	\bibitem [{\citenamefont {Tanaka}(1993{\natexlab{c}})}]{tanaka93-3}%
	\BibitemOpen
	\bibfield  {author} {\bibinfo {author} {\bibfnamefont {H.}~\bibnamefont
			{Tanaka}},\ }\href {https://doi.org/10.1209/0295-5075/24/8/008} {\bibfield
		{journal} {\bibinfo  {journal} {Europhysics Letters ({EPL})}\ }\textbf
		{\bibinfo {volume} {24}},\ \bibinfo {pages} {665} (\bibinfo {year}
		{1993}{\natexlab{c}})}\BibitemShut {NoStop}%
	\bibitem [{\citenamefont {Keblinski}\ \emph {et~al.}(1993)\citenamefont
		{Keblinski}, \citenamefont {Ma}, \citenamefont {Maritan}, \citenamefont
		{Koplik},\ and\ \citenamefont {Banavar}}]{PWA93}%
	\BibitemOpen
	\bibfield  {author} {\bibinfo {author} {\bibfnamefont {P.}~\bibnamefont
			{Keblinski}}, \bibinfo {author} {\bibfnamefont {W.-J.}\ \bibnamefont {Ma}},
		\bibinfo {author} {\bibfnamefont {A.}~\bibnamefont {Maritan}}, \bibinfo
		{author} {\bibfnamefont {J.}~\bibnamefont {Koplik}},\ and\ \bibinfo {author}
		{\bibfnamefont {J.~R.}\ \bibnamefont {Banavar}},\ }\href
	{https://doi.org/10.1103/PhysRevE.47.R2265} {\bibfield  {journal} {\bibinfo
			{journal} {Phys. Rev. E}\ }\textbf {\bibinfo {volume} {47}},\ \bibinfo
		{pages} {R2265} (\bibinfo {year} {1993})}\BibitemShut {NoStop}%
	\bibitem [{\citenamefont {Troian}(1994)}]{troian94}%
	\BibitemOpen
	\bibfield  {author} {\bibinfo {author} {\bibfnamefont {S.~M.}\ \bibnamefont
			{Troian}},\ }\href {https://doi.org/10.1103/PhysRevLett.72.3739} {\bibfield
		{journal} {\bibinfo  {journal} {Phys. Rev. Lett.}\ }\textbf {\bibinfo
			{volume} {72}},\ \bibinfo {pages} {3739} (\bibinfo {year}
		{1994})}\BibitemShut {NoStop}%
	\bibitem [{\citenamefont {Jaiswal}\ \emph
		{et~al.}(2012{\natexlab{b}})\citenamefont {Jaiswal}, \citenamefont {Puri},\
		and\ \citenamefont {Das}}]{PSS12-epl}%
	\BibitemOpen
	\bibfield  {author} {\bibinfo {author} {\bibfnamefont {P.~K.}\ \bibnamefont
			{Jaiswal}}, \bibinfo {author} {\bibfnamefont {S.}~\bibnamefont {Puri}},\ and\
		\bibinfo {author} {\bibfnamefont {S.~K.}\ \bibnamefont {Das}},\ }\href
	{https://doi.org/10.1209/0295-5075/97/16005} {\bibfield  {journal} {\bibinfo
			{journal} {{Europhysics Letters} (EPL)}\ }\textbf {\bibinfo {volume} {97}},\
		\bibinfo {pages} {16005} (\bibinfo {year} {2012}{\natexlab{b}})}\BibitemShut
	{NoStop}%
\end{thebibliography}
\newcommand{\noopsort}[1]{} \newcommand{\printfirst}[2]{#1}
\newcommand{\singleletter}[1]{#1} \newcommand{\switchargs}[2]{#2#1}

\end{document}

\title{SUPPLEMENTAL MATERIAL \\ Universal Fast Mode and Potential-dependent Regimes in Wetting Kinetics
}

%\author{Syed Shuja Hasan Zaidi$^1$, Prabhat K. Jaiswal$^1$, Madhu Priya$^2$, and Sanjay Puri$^3$}
%\email[Corresponding author:\; ]{}
\author{Syed Shuja Hasan Zaidi}
\author{Prabhat K. Jaiswal}
%\email[]{prabhat.jaiswal@iitj.ac.in}
\affiliation{Department of Physics, Indian Institute of Technology Jodhpur, Karwar 342030, India}
\author{Madhu Priya}
%\email[]{madhupriya@bitmesra.ac.in}
\affiliation{Department of Physics, Birla Institute of Technology Mesra, Ranchi  835215, India}
\author{Sanjay Puri}
%\email[]{puri@mail.jnu.ac.in}
\affiliation{School of Physical Sciences, Jawaharlal Nehru University, New Delhi 110067, India}

%\affiliation{$^1$Department of Physics, Indian Institute of Technology Jodhpur, Karwar 342030, India}
%\affiliation{$^2$Department of Physics, Birla Institute of Technology Mesra, Ranchi  835215, India}
%\affiliation{$^3$School of Physical Sciences, Jawaharlal Nehru University, New Delhi 110067, India}

\date{\today}

\maketitle

\section{Methodology}
\par The particle-particle interaction potential is taken to be $12-6$ LJ potential:
\begin{equation}\label{eq4:LJ}
%\begin{split}
	\phi(r) = 4\epsilon_{\alpha\beta}\left[ \left(\frac{\sigma}{r}\right)^{12}- \left(\frac{\sigma}{r}\right)^6\right] .
%V(r) & =\phi(r)-\phi(R_c)-(r-R_c)\frac{d\phi}{dr}\bigg|_{r=R_c} 
%\end{split}
\end{equation}
Here, $r$ is the inter-particle separation and the LJ interaction parameters are set as $\epsilon_{AA} =\epsilon_{BB}=2\epsilon_{AB}=\epsilon$. The potential is truncated, shifted, and force-corrected at the cutoff, $R_c=2.5~\sigma$ \cite{MD17}. We will use LJ reduced units, where the variables are expressed in terms of $\sigma$, $\epsilon$, and $m$. The particle masses are chosen to be equal, $m_A = m_B = m$. Further, the values of $\sigma$, $\epsilon$, and $m$ are set to $1$, and the reduced LJ time unit is $t_0=\sqrt{m\sigma^2/\epsilon} = 1$.

\par
Next, we describe the protocol used in our MD simulation of SDSD. We start with the random placement of $ N=N_A+N_B $ particles in the simulation box, such that the bulk density is 1.  We equilibrate the system at a high temperature with periodic boundary conditions in all directions. At $t=0$, the system is quenched to $T=1\approx 0.7 \,T_c$ (bulk $ T_c\approx 1.423$), and the walls are introduced at $ z=0, D $. The introduction of the walls at $t=0$ ensures that there are no surface-induced inhomogeneities for $t<0$. The Nos\'e-Hoover thermostat is used to control the temperature of the system at all times. Nos\'e-Hoover preserves the relevant features of hydrodynamics in domain growth \cite{SSS10,SSS12,allen96}. Newton’s equations of motion are integrated numerically using the velocity Verlet algorithm \cite{allen96}, with a time step $\Delta t = 0.01$ in LJ units.

\section{Theory}
\subsection*{ The phenomenological theory of the growth of wetting layer:}
Let us briefly discuss the growth exponent seen in \textbf{Fig. 4} of the main text, using the phenomenological model on SDSD \cite{SK01, puri05}. We start by considering the typical depth profiles in \textbf{Fig. 2}. The current in the $z$-direction which drives the wetting-layer growth can be expressed as,
\begin{equation}
\label{eq5:J}
J_z \simeq -\,\frac{dV(z)}{dz}{\bigg\rvert}_{z=R_1} - \frac{\gamma}{Lh}.
\end{equation}
Here, the first term on the right-hand side describes the growth of the wetting layer due to the surface potential gradient. The second term corresponds to the chemical-potential difference between the flat wetting layer and the $A$-rich bulk tubes, assuming the separation between the two entities is $h(t) = R_2(t) - R_1(t)$ (measures the depletion-layer thickness). The variable $h(t)$ can be further approximated as $h(t) \simeq [(1-\psi_0)/(1+\psi_0)]R_1(t)$
by assuming that the wetting layer and depletion layers together constitute an overall composition of $\psi_0$. 

\par Now, for the long-ranged surface fields of the form $V(z) = -\epsilon_a/z^n$, we have 
\begin{equation} \label{eq7:Jz}
J_z \simeq -\frac{n\,\epsilon_a}{R_1^{n+1}} - \frac{\gamma}{LR_1}\left(\frac{1+\psi_0}{1-\psi_0} \right).
\end{equation}
As $J_z = -dR_1/dt$, we finally get the following growth regimes \cite{puri05, SK01},
\begin{equation}\label{eq8:R1}
R_1(t) \sim
\begin{dcases}
(\epsilon_a ~ t)^{1/(n+2)}, & t \ll t_c \, ,\\
g(\psi_0) ~ (\gamma ~ t)^{1/3}, & t \gg t_c\, .
\end{dcases}
\end{equation}
where, the definition of $g(\psi_0)$ can be found in Ref. \cite{SK01}. The crossover time $t_c$ can be obtained by equating the early-time and late-time length scales. From the above equation, it is evident that the potential-dependent regime [$R_1(t) \sim t^{1/(n+2)}$] depends on the surface field strength $\epsilon_a$ as well as the exponent $n$.

\newpage
\section{Figures}
\subsection*{Formation of Connecting tubes from the bulk to the wetting layer during fast~mode and  viscous hydrodynamic regime}

\renewcommand{\figurename}{Fig.}
\renewcommand{\thefigure}{S1}
\begin{figure}[h]
	\centering
	\includegraphics[width=1.0\textwidth]{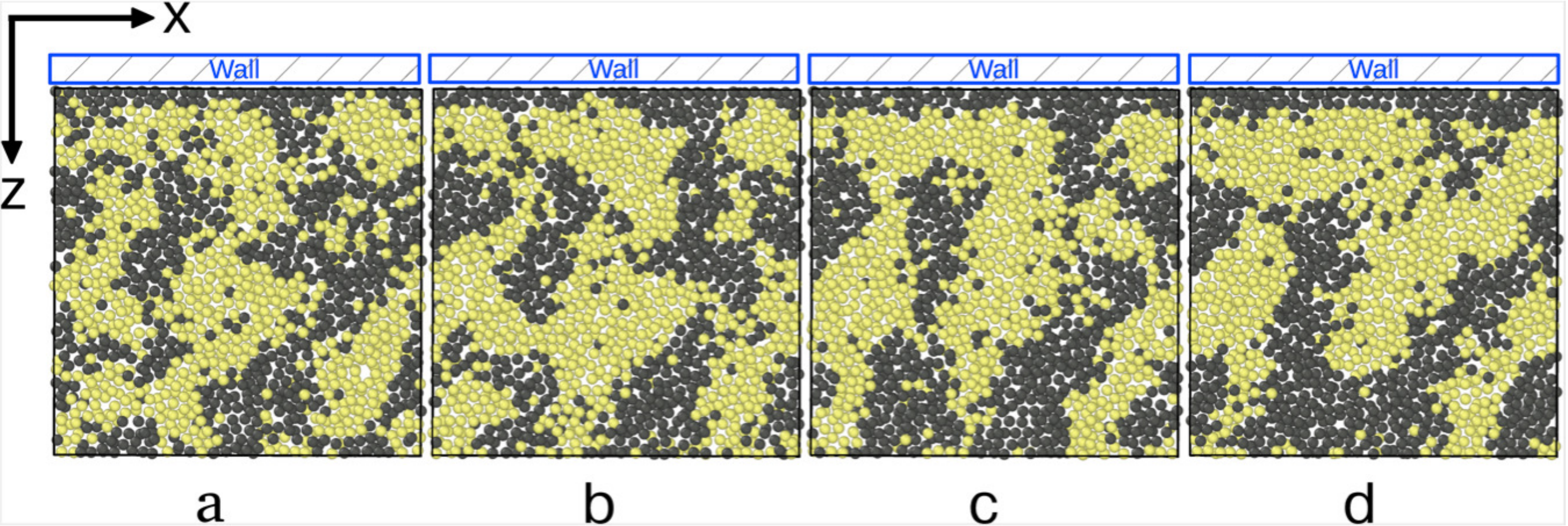}
	\caption{ The bulk tube building up the wetting layer at $t=200$ \textbf{{\fontfamily{phv}\selectfont (a)}} during the diffusive regime, attaches itself to it for sometime, $t=300$ \textbf{{\fontfamily{phv}\selectfont (b)}} and $t=400$ \textbf{{\fontfamily{phv}\selectfont (c)}}, flushing the particles at the surface and disconnecting from the bulk before $t=500$ \textbf{{\fontfamily{phv}\selectfont (d)}}. These snapshots (and movie SM1.avi) correspond to viscous hydrodynamic regime where the wetting layer grows into the bulk.}
	\label{fig1}
\end{figure}

\renewcommand{\figurename}{Fig.}
\renewcommand{\thefigure}{S2}
\begin{figure}[h]
	\centering
	\includegraphics[width=1.0\textwidth]{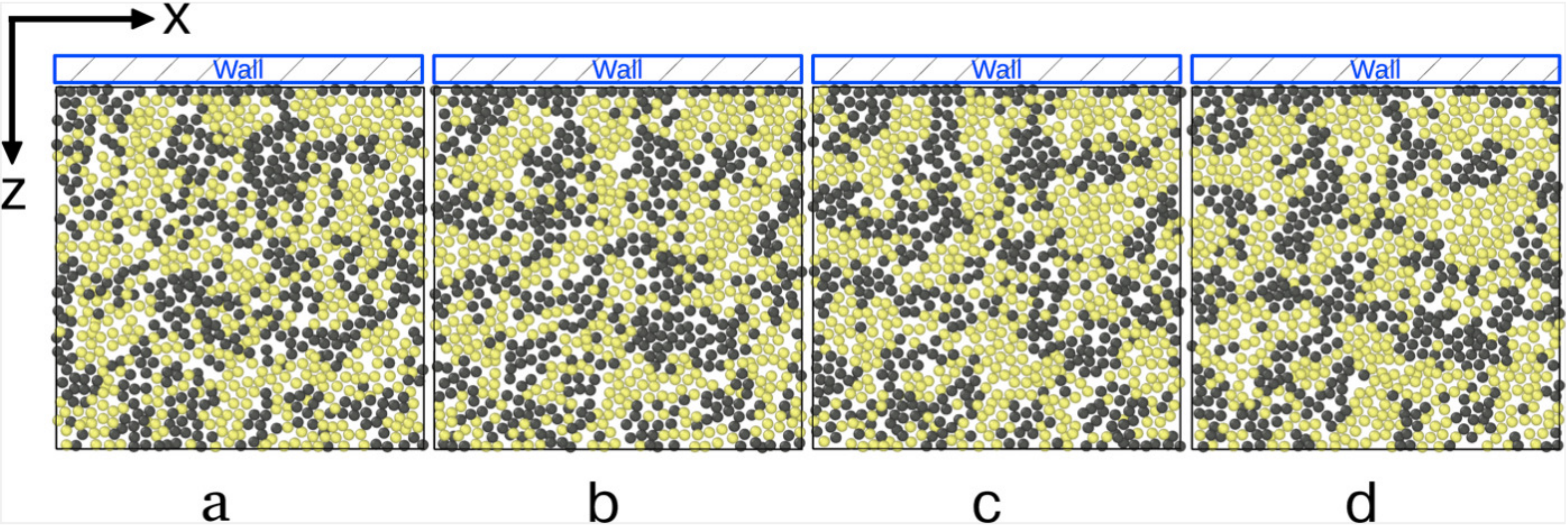}
	\caption{%Vertical bulk tubes approach the surface with no considerable wetting layer yet formed at $t=70$ \textbf{{\fontfamily{phv}\selectfont (a)}}, remains connected for a while, $t=90$ \textbf{{\fontfamily{phv}\selectfont (b)}} and $t=95$ \textbf{{\fontfamily{phv}\selectfont (c)}}, pours down particles onto the wetting layer and then disconnects around $t=120$ \textbf{{\fontfamily{phv}\selectfont (d)}}.
		Vertical bulk tubes approach the surface with no considerable wetting layer yet formed at $t=70$ \textbf{{\fontfamily{phv}\selectfont (a)}} and remain connected for a while, $t=90$ \textbf{{\fontfamily{phv}\selectfont (b)}} and $t=95$ \textbf{{\fontfamily{phv}\selectfont (c)}} , drain the prefered particles onto the surface droplets, and then disconnect around $t=120$ \textbf{{\fontfamily{phv}\selectfont (d)}}. These snapshots (and movie SM2.avi) clearly demosntrate that the fast mode is associated with the \emph{coating dynamics}.}
	\label{fig2}
\end{figure}

In addition, the respective animations of Fig~\ref*{fig1} and \ref*{fig2}, which renders the growth mechanism corresponding to the fast mode and the viscous hydrodynamic regime are available on the website page.

\section{Movies}

\begin{enumerate}
	\item[\textbf{SM1:}]Movie for the viscous hydrodynamic regime for the LJ time 200-500. The snapshots for Fig.~\ref{fig1} are taken from this movie.  
	\item[\textbf{SM2:}] Movie for fast mode for the range of LJ time 70-300. The snapshots for Fig.~\ref{fig2} belong to this movie. 
\end{enumerate}

\section{Results for short-ranged potential}
\subsection*{The results and discussion for short-ranged exponential surface potential}
\par We also simulate the present system with an exponential wall potential $V (z') = -\epsilon_a exp(-z'/z_0 )$ [ see Eq.~2 of the main text ] for the surface at $z=0$. We focus on the growth of the wetting layer with special emphasis on early-stage kinetics. Figure~\ref{fig3} shows the evolution of $R_1(t)$ averaged over 400 independent runs for two different surface field strengths. The early-time regime is short but our MD data is consistent with logarithmic growth $R_1(t)\sim z_0 \ln(\epsilon_a t/z_0^2)$, as predicted earlier \cite{puri05}. This is highlighted in the inset of Fig.~\ref{fig3}, where we plot $R_1$ vs. $t$ on a log-linear scale. The main figure plots $R_1$ vs. $t$ on a log-log scale, which shows the late-time behavior analogous to the long-ranged case. The wetting kinetics exhibits the usual power-law growth ($R_1(t) \sim t^{\alpha}$) with different exponents $\alpha$ corresponding to the respective regimes. The universal diffusive regime ($\alpha=1/3$) is preceded by a transient faster growth $(\alpha \approx 3/2)$ associated with the fast-mode kinetics. We conclude that the fast-mode kinetics is \emph{universal} in nature. Additionally, at very late times, we observe the well-known viscous hydrodynamic regime with an exponent $\alpha\approx 1$.

\renewcommand{\figurename}{Fig.}
\renewcommand{\thefigure}{S3}
\begin{figure}[!ht]
	\centering
	\includegraphics[width=0.5\textwidth]{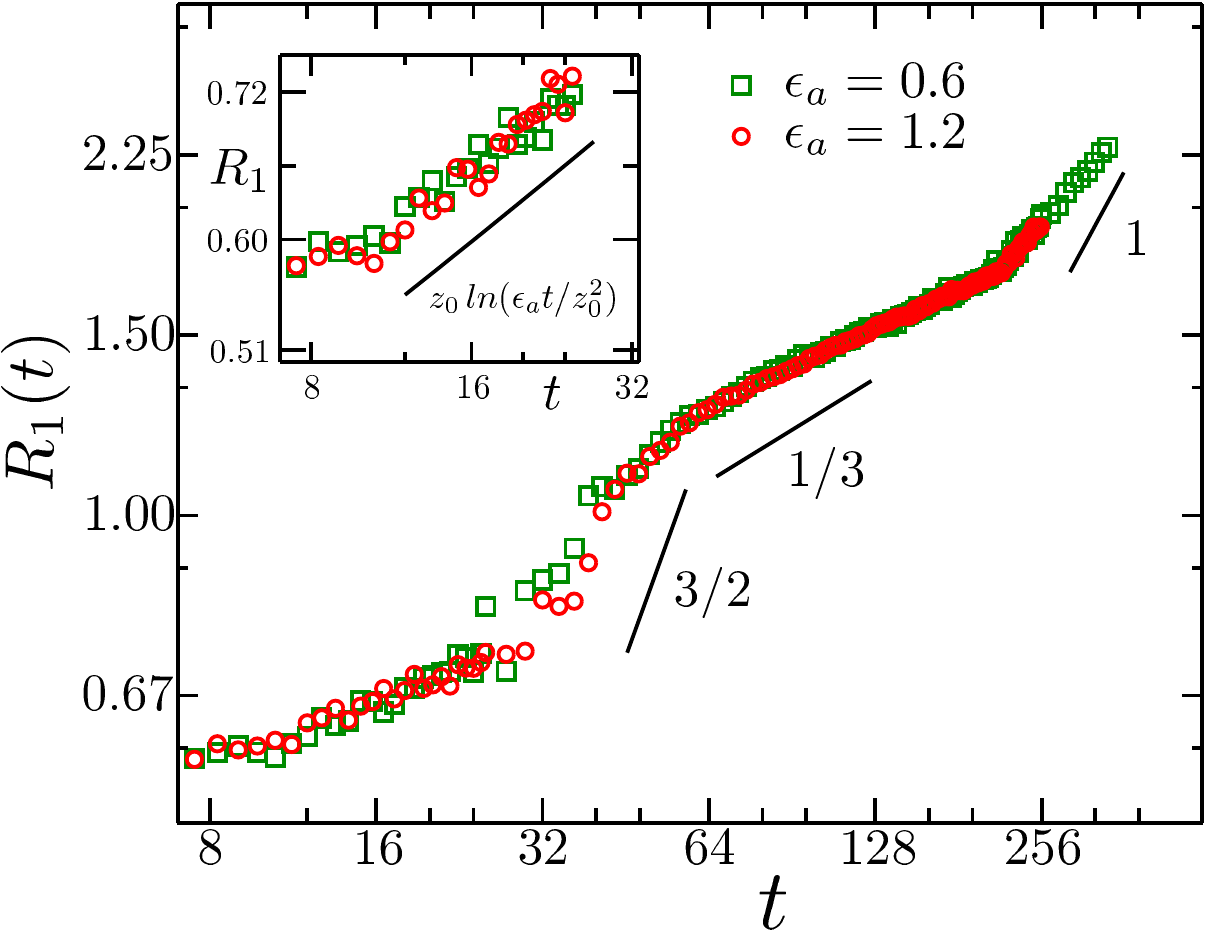}
	\caption{ The wetting-layer thickness $R_1(t)$ with time $t$ on a log-log scale for a short-ranged surface field $V (z) = -\epsilon_a \exp(-z/z_0 )$. The parameter $z_0$ is set to 0.5, and the field strengths are $\epsilon_a=0.6$ (green square) and $\epsilon_a=1.2$ (red circle). It is significant to see the fast mode in this case too. At late times, the plot exhibits the diffusive and hydrodynamic regimes, in succession. Inset: Early time behavior of $R_1(t)$ illustrating a logarithmic growth. The straight line has a slope $z_0 \ln(\epsilon_a t / z_0^2)$.}
	%Main figure: $R_1(t)$ for four different surface field strengths $\epsilon_a$ in $V(z)$, shown in legend of the graph. The solid line describes the exponential growth for $h_1=2.4$. Inset: all the curves of lower surface field strength have been collapsed on $h_1=2.4$.}
	\label{fig3}
\end{figure}

\newpage

\bibliographystyle{apsrev4-2}
%\bibliography{supp.bib}
 \newcommand{\noopsort}[1]{} \newcommand{\printfirst}[2]{#1}
 \newcommand{\singleletter}[1]{#1} \newcommand{\switchargs}[2]{#2#1}